\documentclass[english]{IEEEtran}
\usepackage[T1]{fontenc}
\usepackage[latin9]{inputenc}
\usepackage{babel}
\usepackage{array}
\usepackage{float}
\usepackage{units}
\usepackage{multirow}
\usepackage{amsmath}
\usepackage{amssymb}
\usepackage{mathdots}
\usepackage{mathtools}
\usepackage{graphicx}
\usepackage[unicode=true]{hyperref}

\binoppenalty=\maxdimen
\relpenalty=\maxdimen

\makeatletter

\providecommand{\tabularnewline}{\\}
\floatstyle{ruled}
\newfloat{algorithm}{tbp}{loa}
\providecommand{\algorithmname}{Algorithm}
\floatname{algorithm}{\protect\algorithmname}

\usepackage{stmaryrd}
\usepackage{blkarray}

\@ifundefined{showcaptionsetup}{}{%
 \PassOptionsToPackage{caption=false}{subfig}}
\usepackage{subfig}
\makeatother

\begin{document}
\title{Natural Steganography in JPEG Domain \\ with a Linear Development
Pipeline}
\author{Théo Taburet$^\times$, Patrick Bas$^\times$, Wadih Sawaya$^\sharp$,
and Jessica Fridrich$^+$\\$^\times$CNRS, Ecole Centrale de Lille,
CRIStAL Lab, 59651 Villeneuve d\textquoteright Ascq Cedex, France,
\{theo.taburet,patrick.bas\}@centralelille.fr\\$^\sharp$IMT Lille-Douai,
Univ. Lille, CNRS, Centrale Lille, UMR 9189, France, wadih.sawaya@imt-lille-douai.fr\\$^+$Department
of ECE, SUNY Binghamton, NY, USA, fridrich@binghamton.edu}
\maketitle

\IEEEpeerreviewmaketitle

\begin{abstract}
In order to achieve high practical security, Natural Steganography (NS) uses cover images captured at ISO sensitivity $ISO_{1}$ and generates stego images mimicking ISO sensitivity $ISO_{2}>ISO_{1}$.
This is achieved by adding a stego signal to the cover that mimics the sensor photonic noise. This paper proposes an embedding mechanism to perform NS in the JPEG domain after linear developments by explicitly computing the correlations between DCT coefficients before quantization.
In order to compute the covariance matrix of the photonic noise in the DCT domain, we first develop the matrix representation of demosaicking, luminance averaging, pixel section, and 2D-DCT. A detailed analysis of the resulting covariance matrix is done in order to explain the origins of the correlations between the coefficients of $3\times3$ DCT blocks. 
An embedding scheme is then presented that takes into account all the correlations. It employs 4 sub-lattices and 64 lattices per sub-lattices. The modification probabilities of each DCT coefficient are then derived by computing conditional probabilities computed from a multivariate Gaussian distribution using the Cholesky decomposition of the covariance matrix. This derivation is also used to compute the embedding capacity of each image. Using a specific database called {\it E1Base}, we show that in the JPEG domain NS (J-Cov-NS) enables to achieve high capacity (more than 2 bits per non-zero AC DCT) and with high practical security ($P_{\mathrm{E}}\simeq40\%$ using DCTR and $P_{\mathrm{E}}\simeq32\%$ using SRNet) from QF 75 to QF 100).
\end{abstract}
\section{Introduction}

In 1998, Cachin~\cite{Cachin:1998:ITM} defined the theoretical security
of a steganographic embedding scheme as $D_{KL}(P_{X},P_{Y})$, the
Kullback--Leibler divergence between the distributions of the cover
contents $P_{X}$ and stego contents $P_{Y}$. Using this definition,
a scheme providing $D_{KL}(P_{X},P_{Y})=0$ should be theoretically
perfectly secure.

Interestingly, only few exceptions, such as Model-Based Steganography
(MBS)~\cite{Sallee03}, HUGO~\cite{ih10Pevny}, and MiPOD~\cite{sedighi2016content},
are based on Cachin's rationale, while the majority of embedding schemes,
such as UNIWARD~\cite{holub2014universal}, HILL~\cite{li2014new},
and UERD~\cite{guo2015using} minimize the sum of empirically defined
costs based on the local complexity of each pixel/DCT coefficient.
In MBS, the embedding preserves the underlying generalized Cauchy
distribution fit to each DCT mode. In HUGO, the cost is computed from
the difference between the SPAM features set~\cite{pevnySPAM} used
for steganalysis. MiPOD minimizes the deflection coefficient, i.e.,
the normalized difference between the expectations of the likelihood
ratio under the two hypotheses in the weak signal and large data sample
asymptotics, as a ``cost.''

Natural Steganography (NS)~\cite{bas:WIFS-16,bas:ICASSP-17,denemark:hal-01687194,taburet:2019}
is based on the same principle as model based steganography since
it embeds message whose associated stego signal tries to mimic the
statistical properties of the camera photonic noise, a.k.a. camera
shot noise. Starting with a cover image acquired at $ISO_{1}$, the
embedding is designed in such a way that the stego image looks like
an image acquired at a larger ISO sensitivity $ISO_{2}>ISO_{1}$.
This strategy is named ``cover-source switching'' since it relies
on changing the model of the cover-source during the embedding process.
In the pixel domain or for monochrome sensors~\cite{bas:WIFS-16,bas:ICASSP-17,denemark:hal-01687194,taburet:2019},
this approach has been shown to achieve both high capacity and statistical
undetectability as long as the embedder is able to correctly model
the added signal. The high security of NS schemes is also due to the
fact that NS uses a pre-cover at the embedder~\cite{bas:WIFS-16}.
In contrast to other schemes relying on side information, such as
SI-UNIWARD~\cite{holub2014universal} or other side-informed implementations~\cite{denemark2015side},
the embedding capacity of NS is only limited by the gap between the
two ISO sensitivities.

In the spatial domain, implementations of NS have been proposed for
monochrome sensors, which do not perform demosaicking, with a development
processes that includes only quantization, gamma correction~\cite{bas:WIFS-16},
and downsampling~\cite{bas:ICASSP-17}. In the JPEG domain, previous
works~\cite{denemark:hal-01687194,taburet:2019} have shown
that models that only consider first-order marginal statistics (histograms)
work well for monochrome sensors but the embedding is very detectable
for color sensors since the embedding does not take into account dependencies
due to demosaicking.

Note that like side-informed embedding methods~\cite{fridrich2005perturbed,holub2014universal}, NS uses a pre-cover image, being here the RAW image. This application scenario can be practically motivated by the fact that it is nowadays possible for a user to record his acquisition in RAW format, even on smart-phones~\cite{CameraAPI}.

\begin{figure*}[t]
\begin{centering}
\includegraphics[width=0.8\textwidth]{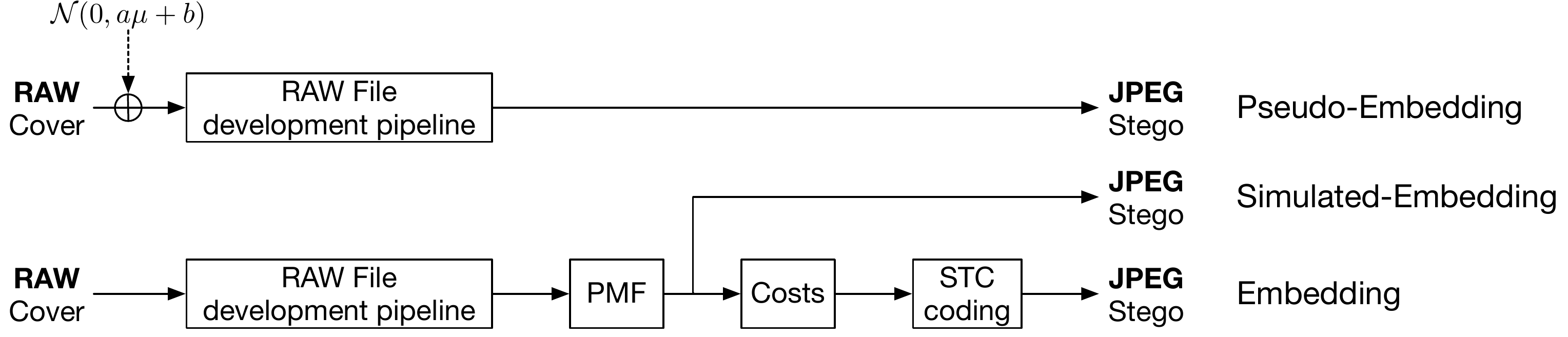}
\par\end{centering}
\medskip{}

\caption{Differences between embedding, simulated embedding, and pseudo embedding.\label{fig:Development-of-a-1}}
\end{figure*}

The goal of this paper is to extend Natural Steganography in the JPEG
domain to color sensors. The paper is organized as follows. Section~\ref{sec:Preliminaries}
introduces notation, and describes the considered development pipeline
and the principle of embedding using NS. Section~\ref{sec:Modeling-dependencies-in}
derives the statistical distribution of the stego signal in the DCT
domain by computing the covariance matrix of its associated joint
distribution. Section~\ref{sec:Analysis-of-the} provides a deep
analysis of different components of the resulting covariance matrix.
Finally, Section~\ref{sec:Embedding-Scheme} presents the embedding
scheme. The new scheme is benchmarked in Section~\ref{sec:Results}
and compared with relevant state-of-the-art steganographic schemes.

This paper is an important extension of the method presented in~\cite{taburet:2019},
where the statistical properties of the photonic noise are obtained
by empirically estimating the noise covariance matrix. The obtained
estimation error leads to a higher detectability, especially for high
JPEG quality factors. In this paper, we instead compute the covariance
matrix exactly as presented in~\cite{taburet:hal-02165866}. We add
an extensive analysis of the properties of this matrix and a detailed
description of the embedding scheme. We also propose a large variety
of results at different JPEG quality factors and for different alphabet
sizes.

\section{Preliminaries\label{sec:Preliminaries}}

\subsection{Notations}

Throughout this article, we use capital letters for random variables $X$ and their corresponding lowercase symbols for their realizations $x$. 
Matrices are written in uppercase $\mathbf{A}$ and vectors (of scalar or random variables) in lowercase boldface font $\mathbf{a}$. Matrix transposition is denoted with a superscript $\mathbf{A}^{t}$.
The subscripts $\bigbox_{p}$ and $\bigbox_{d}$ will be respectively associated to the photo-site domain and the developed domain. 

In this article, matrix vectorization of matrices according to the rows or columns are used. For a $m\times n$ matrix $\mathbf{A}$, the respective vectorization by rows and columns is defined as follows:

For:

\begin{equation}
\mathbf{A}\in\mathbb{R}^{m\times n}/\mathbf{A}=\left[\begin{array}{ccc}
a_{1,1} & \ldots & a_{1,n}\\
\vdots &  & \vdots\\
a_{m,1} & \ldots & a_{m,n}
\end{array}\right]\label{eq:Mat_A_def}
\end{equation}

the respective vectorization by columns $(C)$ and rows $(R)$ is defined as follows:

\begin{equation}
\mathrm{vec_{C}}(\mathbf{A})=\left[a_{1,1},\ldots,a_{m,1},\ldots,a_{1,n},\ldots a_{m,n}\right]^{t}\in\mathbb{R}^{mn\times1}\label{eq:vec_R_def}
\end{equation}

\begin{equation}
\mathrm{vec_{R}}(\mathbf{A})=\left[a_{1,1},\ldots,a_{1,n},\ldots,a_{m,1},\ldots a_{m,n}\right]^{t}\in\mathbb{R}^{mn\times1}\label{eq:vec_C_def}
\end{equation}

\subsection{Pseudo-embedding, simulated embedding and embedding\label{subsec:Pseudo-embedding,-simulated-embe}}

We distinguish between three forms of steganographic embedding that
are illustrated in Figure~\ref{fig:Development-of-a-1}: \emph{pseudo-embedding},
\emph{simulated embedding,} and (true) \emph{embedding}.

\emph{Pseudo-embedding} means that practical embedding is not possible
with the proposed implementation. It acts as a generic mathematical
operation (a reference) which outputs the so-called \emph{pseudo-stego}
image should be statistically distributed like the stego image.

In\emph{ simulated embedding}, the embedding changes are simulated
according to a given selection channel -- the probability $\pi_{i}(k)$
of modifying the cover sample by magnitude $k$ at location $i$.

\emph{(True) embedding} can be realized using multilayered STCs~\cite{filler2011minimizing}
based on costs $\rho_{i}(k)$ directly computed from the set of embedding
probabilities $\pi_{i}(k)$, with $\rho_{i}(k)=\log\left(\pi_{i}(0)/\pi_{i}(k)\right)$.
The STC algorithm minimizes the sum of embedding costs while embedding
the payload using a Viterbi algorithm.

\subsection{Principles of Natural Steganography}

We first review the principles of Natural Steganography when pseudo
embedding is performed at the photo-site level, and then introduce
the technical goals of this paper.

\subsubsection{Pseudo-embedding at the photo-site level}

Modifying the photo-sites directly leads to pseudo-embedding. However,
as mentioned in~\cite{bas:WIFS-16}, it can also be directly used
for simulated embedding or true embedding in the spatial domain for
monochrome sensors.

The key idea here is to add a stego signal $S$ that mimics the statistical
properties of the photonic noise. For a CCD or CMOS sensor, the photonic
noise $N$ at photo-site $i,j$ due to the error of photonics count
during acquisition is assumed to be independent across photo-sites
with a widely adopted heteroscedastic model~\cite{european2010standard}:

\begin{equation}
N_{i,j}^{(1)}\sim\mathcal{N}\left(0,a_{1}\mu_{i,j}+b_{1}\right),\label{eq:Noise_ISO_1}
\end{equation}
where $\mu_{i,j}$ is the noiseless photo-site value at photo-site
$i,j$, and $(a_{1},b_{1})$ a pair of parameters depending only on
the $ISO_{1}$ sensitivity and the specific sensor. The acquired photo-site
sample $x_{i,j}^{(1)}$ is thus a realization $x_{i,j}^{(1)}=\mu_{i,j}+n_{i,j}^{(1)}$
of a Gaussian variable distributed as $X_{i,j}^{(1)}\sim\mathcal{N}\left(\mu_{i,j},a_{1}\mu_{i,j}+b_{1}\right)$.

In the same way, for sensitivity $ISO_{2}$: $X_{i,j}^{(2)}\sim\mathcal{N}\left(\mu_{i,j},a_{2}\mu_{i,j}+b_{2}\right)$.
Thus, we can generate a stego image mimicking a cover captured at
$ISO_{2}$ such that for each photo-site $i,j$ we have:
\begin{equation}
y_{i,j}=x_{i,j}^{(1)}+s_{i,j},\label{eq:pseudo-emb}
\end{equation}
 with $S_{i,j}$ the random variable representing the stego signal:
\begin{equation}
S_{i,j}\sim\mathcal{N}\left(0,\left(a_{2}-a_{1}\right)x_{i,j}+b_{2}-b_{1}\right).\label{eq:Stego_noise_photo}
\end{equation}

The photo-site of the stego image is then distributed as:

\begin{equation}
Y_{i,j}\sim\mathcal{N}\left(\mu_{i,j},a_{1}\mu_{i,j}+b_{1}+\left(a_{2}-a_{1}\right)x_{i,j}+b_{2}-b_{1}\right).\label{eq:Stego_noise_CSS}
\end{equation}
Assuming that the value of the observed photo-site is close to its
expectation, i.e., $\mu_{i,j}\approx x_{i,j}^{(1)}$, we obtain

\begin{equation}
Y_{i,j} \overset{d}{=} X_{i,j}^{(2)},\label{eq:Stego_Noise_ISO_2}
\end{equation}
where $\overset{d}{=}$ represents the equality in distribution of 2 random variables. 
Equation~(\ref{eq:Stego_Noise_ISO_2}) highlights that the distribution
of a stego image photo-site is the same as the distribution of a cover
photo-site acquired at $ISO_{2}$. Equation~(\ref{eq:pseudo-emb})
is the pseudo-embedding operation, which enables us to generate pseudo-stego
content at the photo-site level. Practically, the distribution of
the stego signal in the continuous domain takes into account the statistical
model of the shot noise estimated for two ISO settings, $ISO_{1}$
and $ISO_{2}$, using the procedure described in~\cite{bas:WIFS-16,foi2008practical}.
The work presented in~\cite{bas:WIFS-16,bas:ICASSP-17} shows that
for monochrome sensors, this model in the spatial domain can be used
to derive the distribution of the stego signal in the spatial domain
after quantization, gamma correction, and image downsampling using
bilinear kernels.

\subsubsection{Simulated embedding in JPEG domain}

The main purpose of this paper is to detail how to perform modifications
on quantized DCT coefficients in order to perform simulated embedding.
The modeling of the stego signal and its dependencies in the DCT domain
are crucial for the embedding to be secure. We thus focus on modeling
the image development process in order to firstly derive the statistical
characteristics of the stego signal in the DCT domain, then compute
the modification probabilities for each DCT coefficient, and finally
perform simulated embedding.

The next section, we explain how we reach the first goal and in Section~\ref{sec:Embedding-Scheme}
we detail the algorithm used to perform simulated embedding.

\section{Modeling dependencies in the DCT domain\label{sec:Modeling-dependencies-in}}

\subsection{The development pipeline}

In this paper, we use a linear development pipeline. Since the distribution at the photo-site level of the random vector of components $S_{i,j}$ is multivariate Gaussian (with diagonal covariance matrix), and because the pipeline up to the DCT transform is a succession of linear operations, one main result of statistical signal processing~\cite{blanc1965theory} it that its distribution in the DCT domain is also a multivariate Gaussian distribution, but with arbitrary covariance matrix. The linear development allows us also to derive the covariance matrix of this distribution. We can write that $\mathbf{y}_p=\mathbf{x}_p+\mathbf{s}_p$,
where $\mathbf{x}_p$ is the vectorized version of a block of photo-site
values of the cover image, and $\mathbf{s}_p$ the vectorized values
of the added stego signal in the photo-site domain. 

The goal of this section is to model the development pipeline as a
linear equation in the form of:

\begin{equation}
\mathbf{y}_d=\mathbf{M}\mathbf{y}_p \Leftrightarrow \mathbf{s}_d=\mathbf{M}\mathbf{s}_p,\label{eq:goal}
\end{equation}
where $\mathbf{y}_d$ and $\mathbf{s}_d$ represent the vectors of respectively the stego content and the stego signal in the developed domain. 
%where $\mathbf{y}$ is a vector composed of photo-site values of the stego image as if it has been generated with pseudo-embedding, and $\mathbf{y}_d$ is a vector composed of DCT coefficients before quantization. 

Since the only random component is the stego signal $\mathbf{s}_p$, the covariance matrix $\mathbf{\Sigma}_d=\mathrm{Cov}(\mathbf{s}_d)$ of the multivariate distribution in
the DCT domain will then be given by:

\begin{equation}
\begin{array}{cc}
\mathbf{\Sigma}_d & =\mathbf{M}\,\mathbf{\Sigma}_p\,\mathbf{M}^{t},\end{array}\label{eq:Cov_goal}
\end{equation}
where $\mathbf{\Sigma}_p=\mathrm{Cov}(\mathbf{s}_p)$ is the covariance
matrix of the considered block of the stego signal in the photo-site domain given the cover $\mathbf{x}$. 

Denoting now $i$ the index of one photo-site in $\mathbf{x}_p$, using~(\ref{eq:Stego_noise_photo})
and the hypothesis $\mu_{i}\approx x_{i}$, the covariance
matrix $\mathbf{\Sigma}_p$ of the stego-signal is a
diagonal matrix with diagonal terms equal to $\left(a_{2}-a_{1}\right)x_{i}+b_{2}-b_{1}$.

\begin{figure}[H]
\begin{centering}
\includegraphics[width=0.8\columnwidth]{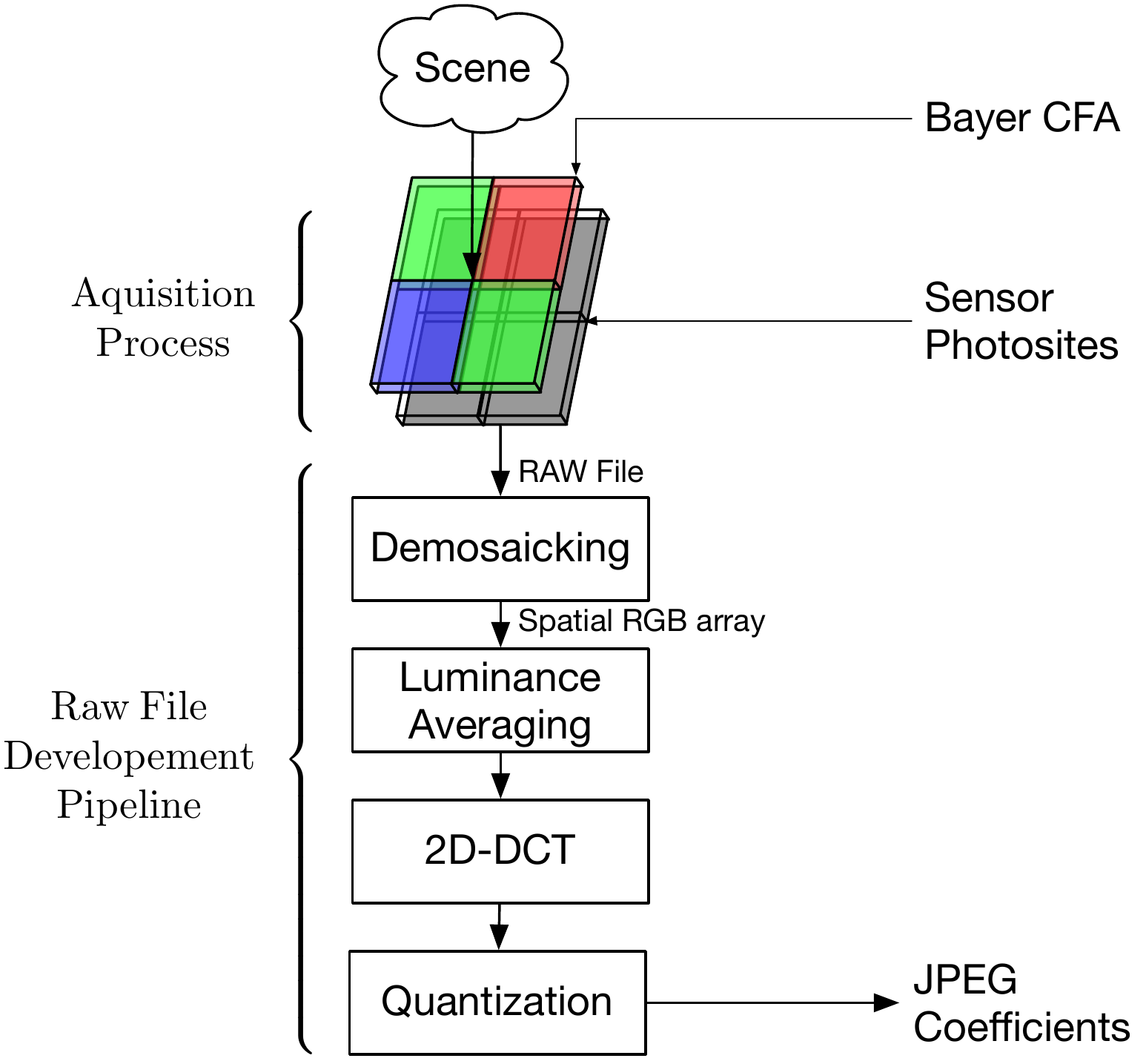}
\par\end{centering}
\caption{Development pipeline: From a scene captured by a color sensor to
luminance JPEG coefficients.\label{fig:Developement-pipeline--2}}
\end{figure}

In order to compute $\mathbf{M}$, we consider the different steps of
the pipeline and decompose the computation of $\mathbf{M}$ into the
following steps (see Figure~\ref{fig:Developement-pipeline--2}):
\begin{enumerate}
\item Demosaicking: this step predicts for each photo-site the two missing
colors that are not recorded by the sensor. We use bilinear filtering
as a linear interpolation process.
\item Luminance averaging: (we only consider embedding in grayscale JPEG image) the demosaicked vector undergoes luminance averaging following the ITU-R BT 601 standard.
\item 2D-DCT transform is computed independently on each block of $8\times8$
pixels.
\item Quantization: the DCT coefficients are quantized using the quantization
table matching a selected JPEG quality factor $(QF)$ to generate
a set of JPEG coefficients. Note that since this operation is non-linear,
it is not captured by equation~(\ref{eq:goal}).
\end{enumerate}
We now detail the different linear operations which are detailed on content vectors $\mathbf{y}_f$ (the subscript $f$ denoting the operation), but can also be written w.r.t. $\mathbf{s}_f$ thanks to the linear formulation by switching $\mathbf{y}_f$ by $\mathbf{s}_f$. 

\subsection{Considered photo-sites}

Since the color interpolation step uses the neighboring photo-sites
to interpolate colors, this creates correlations between adjacent
8-connected blocks of $8\times8$ photo-sites. 

\begin{figure}[h]
\begin{centering}
\includegraphics[width=0.3\columnwidth]{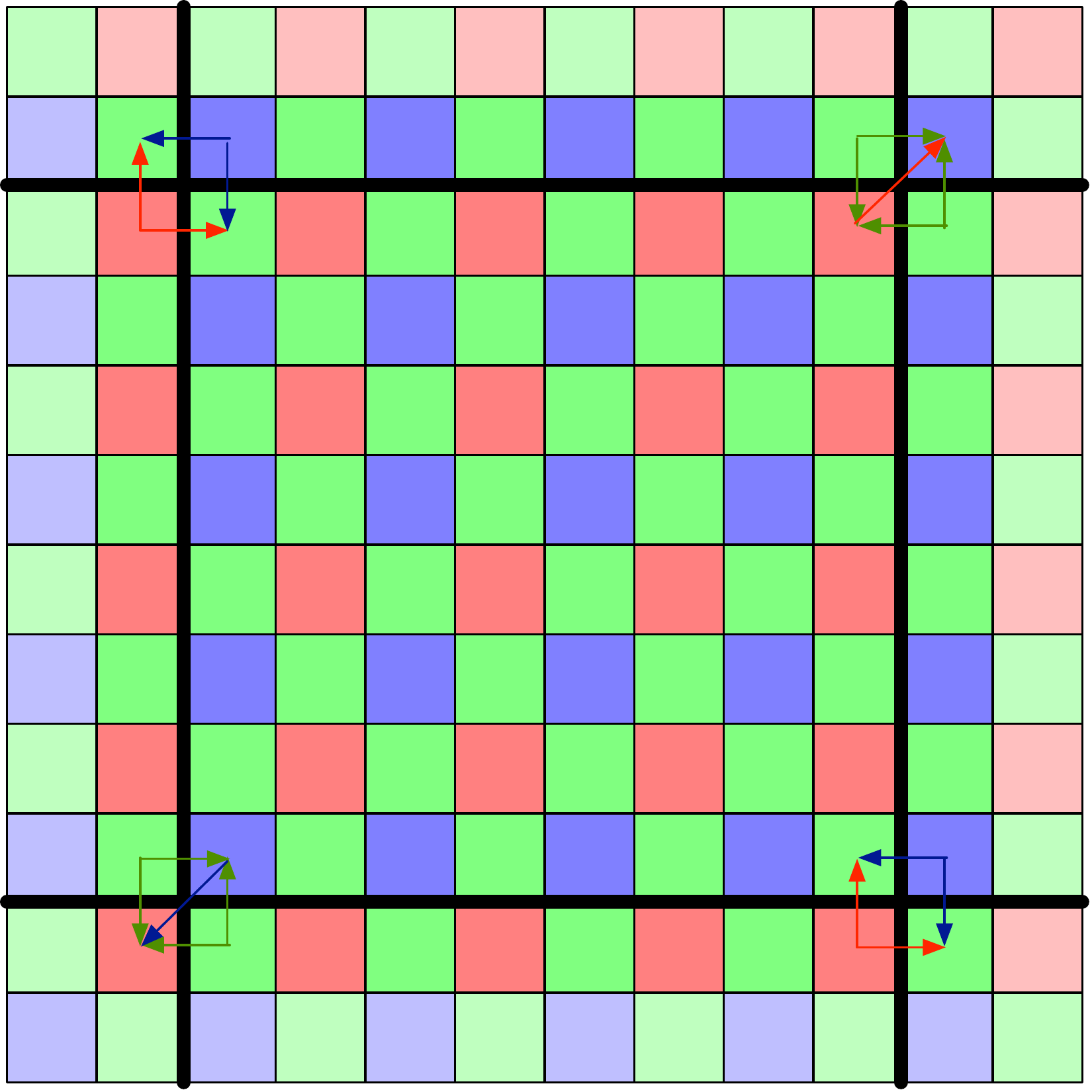}
\par\end{centering}
\medskip{}

\caption{Locations of photo-sites (dark colors) used to interpolate pixel values within one block using bilinear demosaicing. 
Diagonal blocks are involved in the computation on two pixels for the blue channel (up right) and the red channel (bottom, left).\label{fig:Location-of-photo-sites-1}}
\end{figure}

These correlations between blocks can be very weak, especially between diagonal blocks. On the contrary, it is important to note that two blocks which are not 8-connected represent independent realizations of the sensor-noise after demosaicking. This property will be used in Section~(\ref{sec:Embedding-Scheme})
to design the embedding scheme.
Both correlation between adjacent blocks and uncorrelated blocks are illustrated in Figure~\ref{fig:Location-of-photo-sites-1}. On this figure we can see that two diagonal blocks can share only two correlated photo-sites, and the correlations can either come from three photo-site values coming from vertical,
horizontal, and diagonal blocks (this is the case between NE and SW
neighbors), or two photo-site values coming from horizontal and vertical
blocks only (this is the case for NW or SE neighbors). On the contrary two blocks that are disconnected are associated to uncorrelated stego signals. 

\begin{figure}[H]
\begin{centering}
\includegraphics[width=0.75\columnwidth]{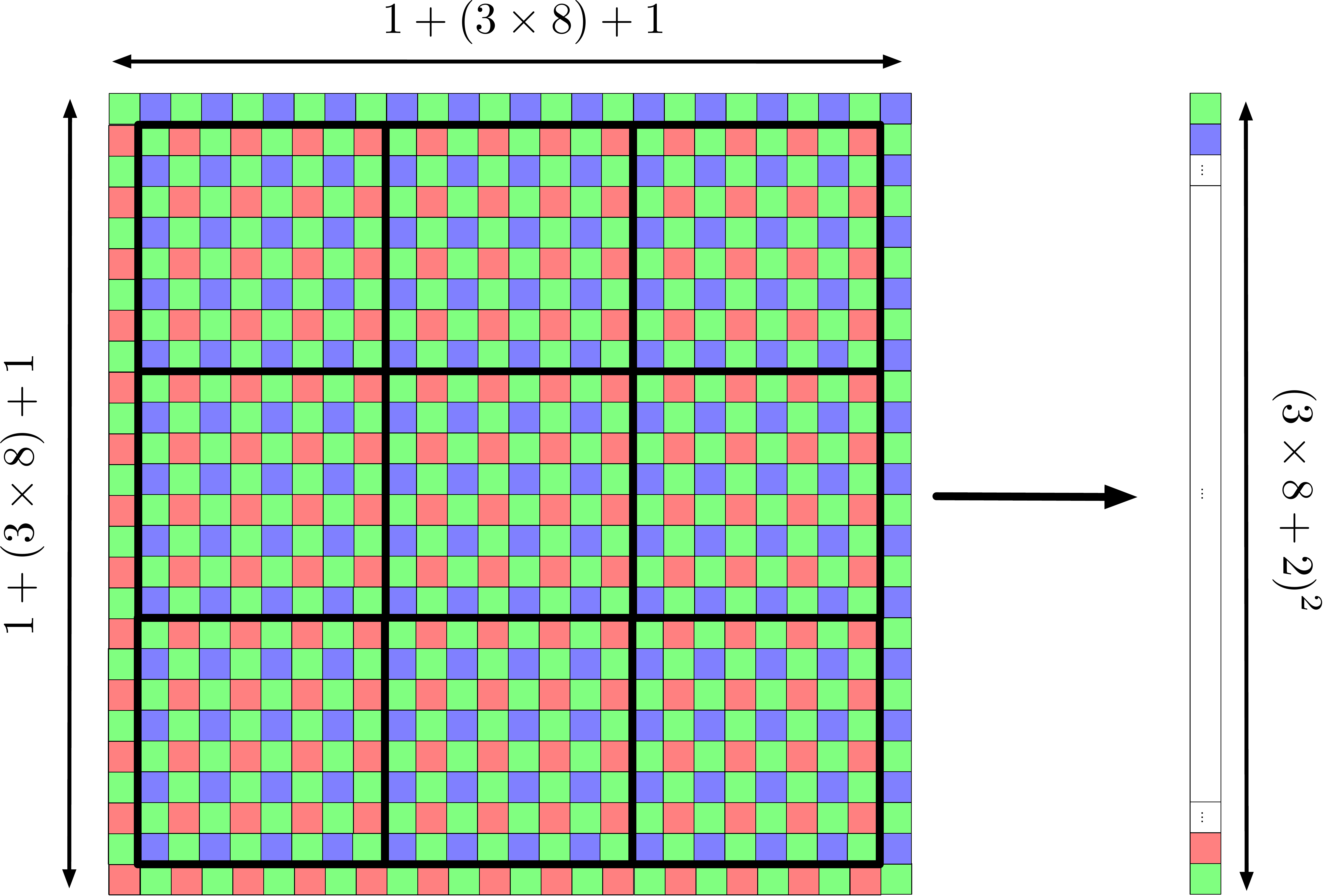}
\par\end{centering}
\caption{RAW photo-sites and its outer border.\label{fig:RAW-photo-sites}}
\end{figure}

In order to capture all the correlations between DCT coefficients,
we consequently need to consider a matrix $\mathbf{Y}_p$ of $(3\times8+2)\times(3\times8+2)$ photo-sites, which gives after vectorization $\mathrm{vec_{R}}(\mathbf{Y}_p)$
a vector $\mathbf{y}_p$ of $676$ photo-sites as an input of our linear
system as illustrated in Figure~\ref{fig:RAW-photo-sites}.

\subsection{Demosaicking}

It is possible to write the demosaicking operations as matrix multiplications.
For each component R, G and B, we define the matrices $\mathbf{D}_{r}$, $\mathbf{D}_{g}$, $\mathbf{D}_{b}$ of size $(24+2)^{2}\times(24+2)^{2}$, such that the result of the matrix multiplication of $\mathbf{y}_p$ with one of these matrices is the vectorized version of the corresponding color channel after
demosaicking:

\begin{equation}
\mathbf{y}_r=\mathbf{D}_r\mathbf{y}_p,\:\mathbf{y}_g=\mathbf{D}_g\mathbf{y}_p,\:\mathbf{y}_b=\mathbf{D}_b\mathbf{y}_p.\label{eq:Vect_Demo}
\end{equation}

Denoting $i$ the index of one photo-site in $\mathbf{y}_i$, one row $i$ of $\mathbf{D}_{k}$, $k\in \{r,g,b\}$ is obtained by vectorization of a $(24+2)\times(24+2)$ matrix with zeros entries except a specific kernel specific kernel $\mathbf{K}_i$ centered on $(i,i)$. This kernel models any kind of interpolation between neighboring photo-sites and $y_i$:
\begin{equation}
\mathrm{row}_{i}(\mathbf{D}_k)=\mathrm{vec}_{R}\left[\begin{array}{cccccc}
\mathbf{0} & \cdots\\
\vdots & \ddots\\
 &  & \ddots\\
 &  &  & \mathbf{K}_i & \mathbf{0} & \cdots\\
 &  &  & \mathbf{0} & \mathbf{0}\\
 &  &  & \vdots &  & \ddots
\end{array}\right].
\end{equation}

Without loss of generality we now focus on the computation of $\mathbf{D}_{g}$, we consequently have two possibilities in this case:

\begin{itemize}
	\item If index $i$ corresponds to a green photo-site on the Bayer CFA, this photo-site does not need color interpolation, i.e.:
	\begin{equation}
		\mathbf{K}_i=[\mathbf{1}].
		\label{eq:Example_G_pixel_no_interpolation}
	\end{equation}
	\item If index $i$ corresponds to a pixel which needs to be interpolated, then:
	\begin{equation}
		\mathbf{K}_i=
		\left[\begin{array}{ccc}
		0.25 & 0 & 0.25\\
		0 & \mathbf{0} & 0\\
		0.25 & 0 & 0.25
		\end{array}\right].
		\label{eq:Example_G_pixel_interpolation}
	\end{equation}
\end{itemize}
The kernel coefficient in bold representing the location $(i,i)$. For the red and blue channels, we use four different convolution kernels $\mathbf{K}_i$ to build $\mathbf{D}_r$ and $\mathbf{D}_b$, which are: 
$$
\left[\mathbf{1}\right],
\left[
\begin{array}{c}
0.5\\
\mathbf{0}\\
0.5
\end{array}
\right],
\left[
\begin{array}{ccc}
0.5 & \mathbf{0} & 0.5
\end{array}
\right]
\mathrm{and}
\left[
\begin{array}{ccc}
0.25 & 0 & 0.25\\
0 & \mathbf{0} & 0\\
0.25 & 0 & 0.25
\end{array}
\right],
$$
and are respectively used for duplication, interpolation between vertical or horizontal photo-sites and interpolation between four diagonal photo-sites.

\subsection{Luminance averaging}

To perform luminance averaging, we can define the matrix $\mathbf{L}$ following the ITU-R BT 601 standard as:

\begin{equation}
\mathbf{y}_l=\underbrace{(0.2126\cdot\mathbf{D}_r + 0.7152\cdot\mathbf{D}_g + 0.0722\cdot\mathbf{D}_b)}_{\mathbf{L}} \, \mathbf{y}_p, 
\label{eq:Vect_lum_averaging}
\end{equation}
with $\mathbf{y}_{l}\text{\ensuremath{\in}}\mathbb{R}^{(24+2)^{2}\times1}$.

\subsection{Pixel selection}

\begin{figure}[H]
\begin{centering}
\includegraphics[width=0.3\columnwidth]{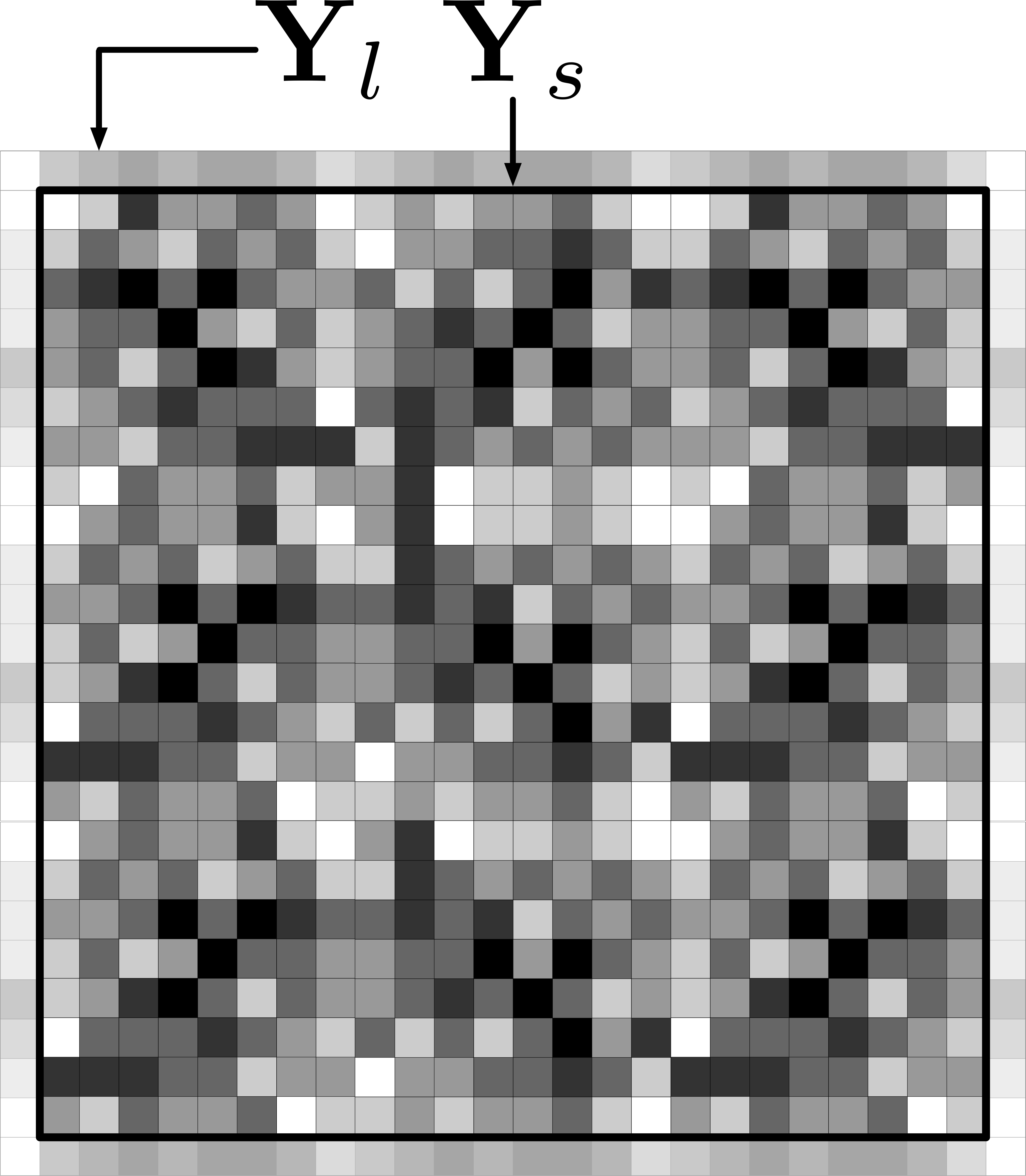}
\par\end{centering}
\caption{Block representation of the pixel selection operation. \label{fig:Scheme-selection}}
\end{figure}

As stated above, the surrounding edges of $3\times3$ blocks of samples are included in order to take into account the convolution window during demosaicking. Once the demosaicking operations have been carried
out, the photo-sites not present in the DCT blocks can then be discarded. Let us denote $\mathbf{Y}_l$ the $(24+2)\times(24+2)$
photo-sites matrix with its outer border, and $\mathbf{Y}_{s}$ without it as depicted in Figure~\ref{fig:Scheme-selection}. The selection matrix $\mathbf{S}\text{\ensuremath{\in}}\mathbb{R}^{(24)^{2}\times(24+2)^{2}}$
can then be defined such that:
\begin{equation}
\mathbf{y}_s=\mathrm{vec_{R}}(\mathbf{Y}_s)=\mathbf{S}\,\mathrm{vec_{R}}(\mathbf{Y}_l)=\mathbf{S}\mathbf{y}_l.
%\Leftrightarrow
%\mathrm{vec_{R}}(\mathbf{L}_s)=\mathbf{\mathbf{S}}\,\mathrm{vec_{R}}(\mathbf{L})
\label{eq:Vect_Selection}
\end{equation}
and we also can write: $\mathbf{y}_s=\mathbf{S}\,\mathbf{L}\mathbf{y}_p$.

\subsection{Blocks permutation and block selection}
Blocks permutation and block selection are not mandatory, but they are useful to compute conditional probabilities while limiting the computational load (see section~\ref{sec:Embedding-Scheme}). Depending on the lattice considered during the embedding (see again section~\ref{sec:Embedding-Scheme}), the correlation matrix can be computed for DCT coefficients belonging to one, five or nine adjacent blocks.

In order to mathematically express a block permutation and selection as the matrix multiplication 
\begin{equation}
\mathbf{y}_{pe}=\mathbf{P}\,\mathbf{y}_{s},
\end{equation} 
we define $\mathbf{Y}_s\text{\ensuremath{\in}}\mathbb{R}^{24\times24}$ as
an array composed of the $3\times3$ blocks of pixels, such that the
vector $\mathbf{y}_{s}=\mathrm{vec_{R}}(\mathrm{\mathbf{Y}_{s}})$, with:

\[
\mathrm{\mathbf{Y}_{s}}=\begin{array}{c}
\left[\begin{array}{ccc}
\fbox{\ensuremath{\mathbf{B_{\mathit{0,0}}}}} & \fbox{\ensuremath{\mathbf{B_{\mathit{0,1}}}}} & \fbox{\ensuremath{\mathbf{B_{\mathit{0,2}}}}}\\
\fbox{\ensuremath{\mathbf{B_{\mathit{1,0}}}}} & \fbox{\ensuremath{\mathbf{B_{\mathit{1,1}}}}} & \fbox{\ensuremath{\mathbf{B_{\mathit{1,2}}}}}\\
\fbox{\ensuremath{\mathbf{B_{\mathit{2,0}}}}} & \fbox{\ensuremath{\mathbf{B_{\mathit{2,1}}}}} & \fbox{\ensuremath{\mathbf{B_{\mathit{2,2}}}}}
\end{array}\right]\end{array},
\]
where $\mathbf{B}_{i,j}\in\mathbb{R}^{8\times8}$ are blocks of $8\times8$
pixels, $0\leq i,\,j\leq2$. We recall that DCT is performed independently on each of these blocks. We need then to extract from $\mathbf{y}_{s}$ the vector corresponding to the required sequence of each block. 

For each block to extract, we define a $8^2\times24^{2}$ block selection matrix $\mathbf{P}_{i,j}$ composed of 3 sub-matrices
$\left[\begin{array}{ccc}
\mathbf{\tilde{P}}_{0} & \mathbf{\tilde{P}}_{1} & \mathbf{\tilde{P}}_{2}\end{array}\right]$, where the size of $\mathbf{\tilde{P}_{i}}$ is $64\times(3 \cdot 64)$, $0\leq i\leq2$.
When extracting $\mathrm{vec_{R}}(\mathbf{B}_{i,j})$, all $\mathbf{\tilde{P}_{k}}$
, $k\neq i$ are set to zero and ${\mathbf{\tilde{P}}_{i}}$ takes the
following entries: 
\[
\mathbf{\tilde{P}}_{i}=\left[\begin{array}{ccccc}
\mathbf{F_{\mathit{j}}} & \mathbf{0} & \mathbf{0} & \cdots & \mathbf{0}\\
\mathbf{0} & \mathbf{F_{\mathit{j}}} & \mathbf{0} & \cdots & \mathbf{0}\\
\mathbf{0} & \mathbf{0} & \mathbf{F_{\mathit{j}}} & \cdots & \mathbf{0}\\
\vdots & \vdots & \mathbf{0} & \ddots & \mathbf{0}\\
\mathbf{0} & \mathbf{0} & \mathbf{0} & \mathbf{0} & \mathbf{F_{\mathit{j}}}
\end{array}\right],
\]
 where $\mathbf{F}_{j}$ is a $8\times24$ sub-matrix consisting of 3 sub-matrices 

 $\left[\begin{array}{ccc}
\mathbf{\tilde{F}}_{0} & \mathbf{\tilde{F}}_{1} & \mathbf{\tilde{F}}_{2}\end{array}\right]$ , each of size $8\times8$. When extracting $\mathrm{vec_{R}}(\mathbf{B}_{i,j})$,
all $\mathbf{\tilde{F}}_{k}$ , $k\neq j$, are set to zero and $\mathbf{\tilde{F}}_{j}=\mathbf{I}_{8}$
, the identity $8\times8$ matrix.

We illustrate this with two examples.

\textit{Example 1:} Suppose we need to extract the vectorized form
of the central block $\mathbf{B}_{1,1}$, i.e., $i=1$ and $j=1.$
We then have:
\[
\mathbf{F}_{1}=\left[\begin{array}{ccc}
\mathbf{0} & \mathbf{I}_{8} & \mathbf{0}\end{array}\right],
\]

and
\[
\mathbf{P}_{1,1}=\left[\begin{array}{cccccccccc}
\mathbf{0} & \cdots & \mathbf{0} & \mathbf{\mathbf{F}}_{1} & \mathbf{0} & \cdots & \mathbf{0} & \mathbf{0} & \cdots & \mathbf{0}\\
\mathbf{0} & \ldots & \mathbf{0} & \mathbf{0} & \mathbf{\mathbf{F}}_{1} &  & \mathbf{0} & \mathbf{0} & \ldots & \mathbf{0}\\
\vdots & \ddots & \mathbf{0} & \vdots & \mathbf{0} & \ddots & \mathbf{0} & \vdots & \ddots & \mathbf{0}\\
\mathbf{0} & \cdots & \mathbf{0} & \mathbf{0} & \mathbf{0} & \mathbf{0} & \mathbf{\mathbf{F}}_{1} & \mathbf{0} & \cdots & \mathbf{0}
\end{array}\right].
\]

The corresponding vector permutation matrix is then 
$$\mathbf{P} = \mathbf{P}_{1,1}.$$

\textit{Example 2:} This additional example is useful for the remaining
of the paper (see Section~\ref{subsec:Decomposition-into-lattices}).
Let us extract from $\mathbf{y}_{s}$ the vector resulting from the
concatenation of the vectorized version of five $8\times8$ blocks
of pixels in a given order, 

\begin{equation*}\label{}
	\resizebox{\columnwidth}{!}{$
	\mathbf{y}_{B}=\left[\mathrm{vec_{R}}(\mathbf{B}_{1,1}),\mathrm{vec_{R}}(\mathbf{B}_{0,0}),\mathrm{vec_{R}}(\mathbf{B}_{0,2}),\mathrm{vec_{R}}(\mathbf{B}_{2,0}),\mathrm{vec_{R}}(\mathbf{B}_{2,2})\right]{}^{t}
	$}
\end{equation*}

%$\mathbf{y}_{B}=\left[\mathrm{vec_{R}}(\mathbf{B}_{1,1}),\mathrm{vec_{R}}(\mathbf{B}_{0,0}),\mathrm{vec_{R}}(\mathbf{B}_{0,2}),\mathrm{vec_{R}}(\mathbf{B}_{2,0}),\mathrm{vect_{R}}(\mathbf{B}_{2,2})\right]{}^{t}$.
The corresponding matrix operation will be:
$$
	\mathbf{P}=\left[\begin{array}{ccccc}
	\mathbf{P}_{1,1} & \mathbf{P}_{0,0} & \mathbf{P}_{0,2} & \mathbf{P}_{2,0} & \mathbf{P}_{2,2}\end{array}\right]^{t}.
$$

\subsection{2D-DCT Transform}

For a $8\times8$ block in the spatial domain, $\mathbf{B}$, its 2D-DCT block version written here as $\mathbf{B}_d$ can be expressed by the following matrix multiplication:

\begin{equation}
	\mathrm{DCT}(\mathbf{B})=\mathbf{A}\cdot\mathbf{B}\cdot\mathbf{A}^{t}=\mathbf{A}\cdot(\mathbf{A}\cdot\mathbf{B}^{t})^{t},\label{eq:Mat_DCT_2D}
\end{equation}
with:

\begin{equation}
	\mathbf{A}=\left[\begin{array}{cccccccc}
	a & a & a & a & a & a & a & a\\
	b & d & e & g & -g & -e & -d & -b\\
	c & f & -f & -c & -c & -f & f & c\\
	d & -g & -b & -e & e & b & g & -d\\
	a & -a & -a & a & a & -a & -a & a\\
	e & -b & g & d & -d & -g & b & -e\\
	f & -c & c & -f & -f & c & -c & f\\
	g & -e & d & -b & b & -d & e & -g
	\end{array}\right],\label{eq:A_DCT_2D}
\end{equation}
and : 
\begin{equation}
	\left[\begin{array}{c}
	a\\
	b\\
	c\\
	d\\
	e\\
	f\\
	g
	\end{array}\right]=\frac{1}{2}\left[\begin{array}{c}
	\cos(\frac{\pi}{4})\\
	\cos(\frac{\pi}{16})\\
	\cos(\frac{\pi}{8})\\
	\cos(\frac{3\pi}{16})\\
	\cos(\frac{5\pi}{16})\\
	\cos(\frac{3\pi}{8})\\
	\cos(\frac{7\pi}{16})
	\end{array}\right].\label{eq:Coeffs_A_DCT_2D}
\end{equation}
It should be observed that the multiplication by $\mathbf{A}$ and
$\mathbf{A}^{t}$ is due to the fact that the DCT transform is separable
and processes the columns and rows independently. In order to compute
the covariance matrix of the spatial signal $\mathbf{B}$, we
use vector notation by transforming the matrix $\mathbf{B}\text{\ensuremath{\in}}\mathbb{R}^{8\times8}$
into a vector $\mathbf{b}\text{\ensuremath{\in}}\mathbb{R}^{64}$
by concatenating the columns. As a result, the $8\times8$ matrix $\mathbf{A}$ is transformed into
a $64\times64$ matrix $\mathbf{A}_v$ given by :
\begin{equation}
\mathbf{A}_v=\left[\begin{array}{cccc}
\mathbf{A} & \mathbf{0} & \ldots & \mathbf{0}\\
\mathbf{0} & \mathbf{A} & \mathbf{0} & \vdots\\
\vdots & \mathbf{0} & \ddots & \mathbf{0}\\
\mathbf{0} & \cdots & \mathbf{0} & \mathbf{A}
\end{array}\right].\label{eq:A_Vect_DCT2_D}
\end{equation}
We also define a transpose operator $\mathbf{T}_r\in\mathbb{R}^{64\times64}$ such as $\mathrm{vec_{c}}(\mathbf{X_{S}^{t}})=\mathbf{T}_r\cdot\mathrm{vec_{c}}(\mathbf{X_{S}})=\mathbf{T}_r\cdot\mathbf{x_{S}}$,
with :

\[
\mathbf{T}_r=(\delta_{r(i),\:c(j)})_{\substack{\mathbf{0}\leq i<64\\
0\leq j<64
}
},
\]
and,

\[
\begin{array}{c}
r(i)=8\left\lfloor \nicefrac{i}{8}\right\rfloor +(i\,\mathrm{mod}(8)),\\
c(j)=8(j\,\mathrm{mod}(8)+\left\lfloor \nicefrac{j}{8}\right\rfloor,
\end{array}
\]
$\delta_{r(i),\:c(j)}$ being the Kronecker function applied to row $r(i)$ and column $c(j)$.

The transpose operation $\mathbf{B}^{t}$ is then equivalent
to the multiplication $\mathbf{T}_r\cdot\mathbf{b}$, and the vector
form of the DCT $8\times8$ block $DCT(\mathbf{B})$ finally becomes:
\begin{equation}
\mathrm{DCT}(\mathbf{b})=\underbrace{\mathbf{A}_v\,\mathbf{T}_r\,\mathbf{A}_v\,\mathbf{T}_r}_{\mathbf{T}_b}\,\mathbf{\mathbf{b}}\label{eq:Vect_DCT_2D_1_blc}
\end{equation}
In order to compute the DCT of $n$ blocks of size $8\times8$ ($n \in\{1,5,9\}$),
we now define :

\[
\mathbf{T}=\left(\begin{array}{ccc}
\mathbf{T}_b &  & \mathbf{0}\\
 & \mathbf{\ddots}\\
\mathbf{0} &  & \mathbf{T}_b
\end{array}\right).
\]
With $\mathbf{T}$ a block diagonal matrix with $n$ matrices $\mathbf{T}_b$ on its diagonal.

\subsection{Whole covariance matrix}

The development pipeline can be then explicitly formulated as 
\begin{equation}
\mathbf{s}_d=\mathbf{M}\mathbf{s}_{p}=\underbrace{\mathbf{T}\,\mathbf{P}\,\mathbf{S}\,\mathbf{L}}_{\mathbf{M}}\,\mathbf{s}_p,\label{eq:Vect_DCT_2D_from_RAW}
\end{equation}

and the covariance matrix is computed as:

% \begin{equation}
% \begin{array}{ccl}
% \mathbf{\Sigma}_d & = &\mathbb{E}\left[\mathbf{s}_d\,\mathbf{s}_d^{t}\right]=\mathbb{E}\left[(\mathbf{y}_d-\mathbf{x}_d)\,(\mathbf{y}_d-\mathbf{x}_d)^{t}\right],\\

%  & = & 
%  \mathbf{M}\,\mathbb{E}\left[\mathbf{s}_p\,\mathbf{s}_p^{t}\right]\,\mathbf{M}^{t},\\
%  & = &\mathbf{M}\,\mathbb{E}\left[(\mathbf{y_p}-\mathbf{x_p})\,(\mathbf{y_p}-\mathbf{x_p})^{t}\right]\,\mathbf{M}^{t}.
% \end{array}\label{eq:Cov_from_RAW}
% \end{equation}
\begin{equation}
\begin{array}{ccl}
\mathbf{\Sigma}_d & =  \mathbf{M}\,\mathbb{E}\left[\mathbf{s}_p\,\mathbf{s}_p^{t}\right]\,\mathbf{M}^{t}.\\
\end{array}\label{eq:Cov_from_RAW}
\end{equation}

Note that for a uniform constant RAW image defined by $\mu=\mathrm{const.}$
(i.e., $\mathbb{E}\left[\mathbf{s}_d\cdot\mathbf{s}_d^{t}\right] \propto \mathbf{I}$),
we obtain $\mathbf{\Sigma}_d \propto\mathbf{M}\mathbf{M^{t}}$. Depending of the number of blocks $n$ considered in the neighborhood ($n \in\{1,5,9\}$, see \ref{subsec:Decomposition-into-lattices}), the size of $\mathbf{\Sigma}_d$ is $(n\times 64,n\times 64)$.

\section{Analysis of the covariance matrix\label{sec:Analysis-of-the}}

In this section, we analyze the properties of the derived covariance matrix and interpret its different components. We show that the inter-block correlations are due to the signal continuity between blocks and that intra-block correlations highlight both artifacts due to demosaicking and due to low-pass filtering.

Note that this analysis is beneficial in order to understand the causes of the observed covariances. This understanding enables %both (i)
to decompose the embedding scheme into independent lattices (see section~\ref{sec:Embedding-Scheme}) but also to pave the road for other synchronization strategies applied to other development pipelines. For example, in~\cite{taburet2020jpeg}, the covariance matrix is limited to the effect of averaging and can be used to synchronize DCT coefficients of classical schemes such as UERD or J-Uniward. In~\cite{li2018defining}, relationships between DCT coefficients to preserve continuities are in line with the presented analysis of the inter-block correlations (see \ref{subsec:Inter-block}).

\begin{figure}[h]
\begin{centering}
\subfloat[]{\begin{centering}
\includegraphics[width=1\columnwidth]{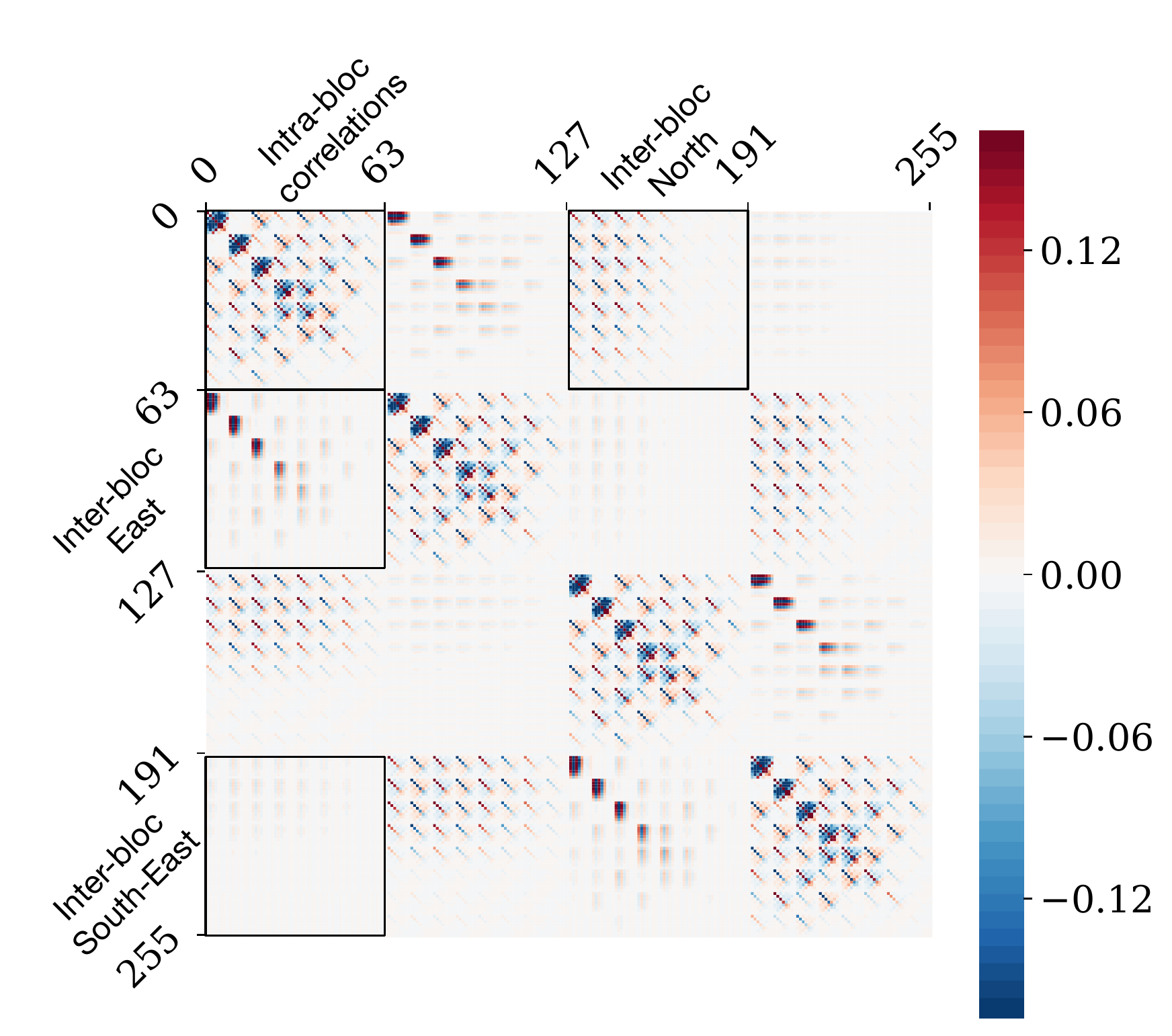}
\par\end{centering}
\label{fig:color}}
\par\end{centering}
\medskip{}

\begin{centering}
\begin{minipage}[c]{0.4\columnwidth}%
\begin{center}
\subfloat[]{\begin{centering}
\includegraphics[width=0.45\columnwidth]{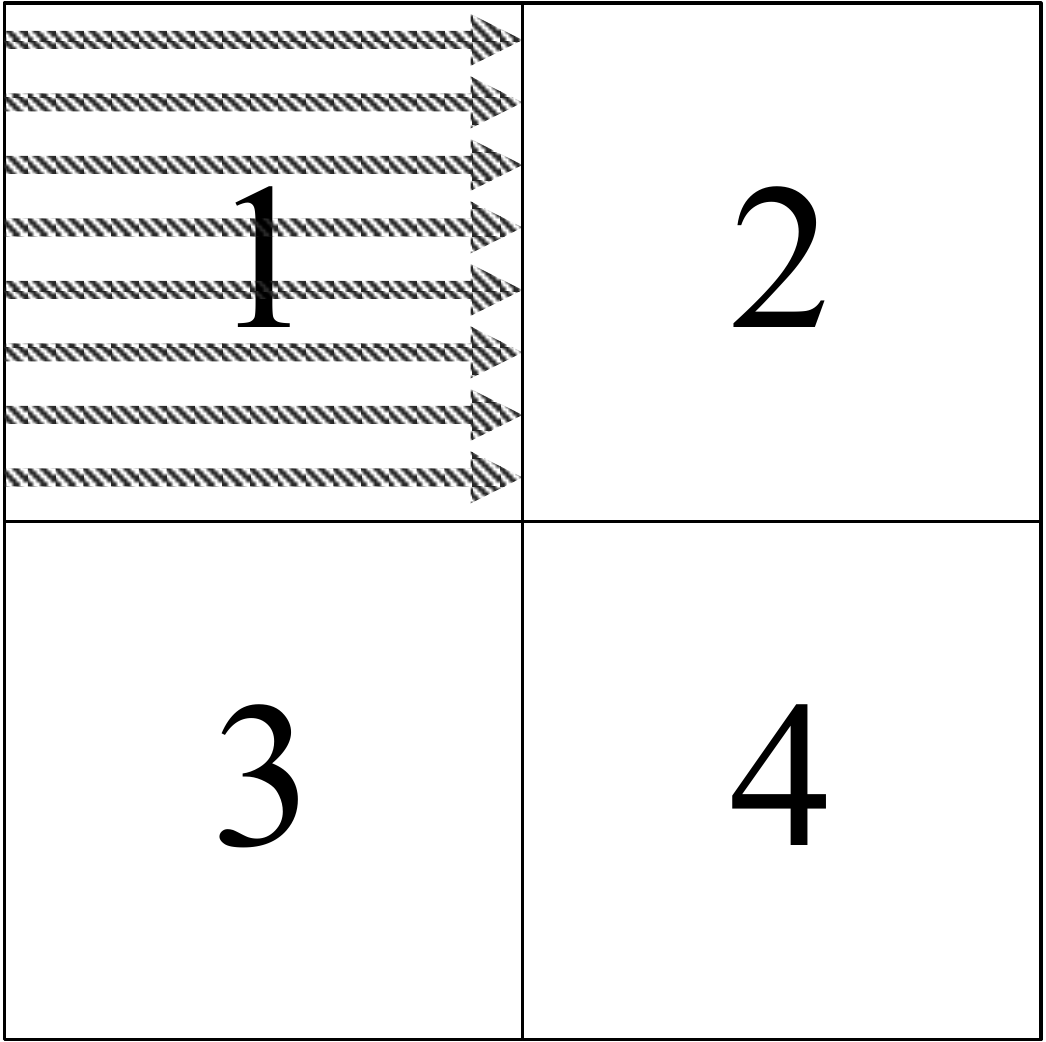}
\par\end{centering}
\label{fig:scan}}
\par\end{center}%
\end{minipage}%
\begin{minipage}[c]{0.5\columnwidth}%
\begin{center}
\subfloat[]{\begin{centering}
\includegraphics[width=0.45\columnwidth]{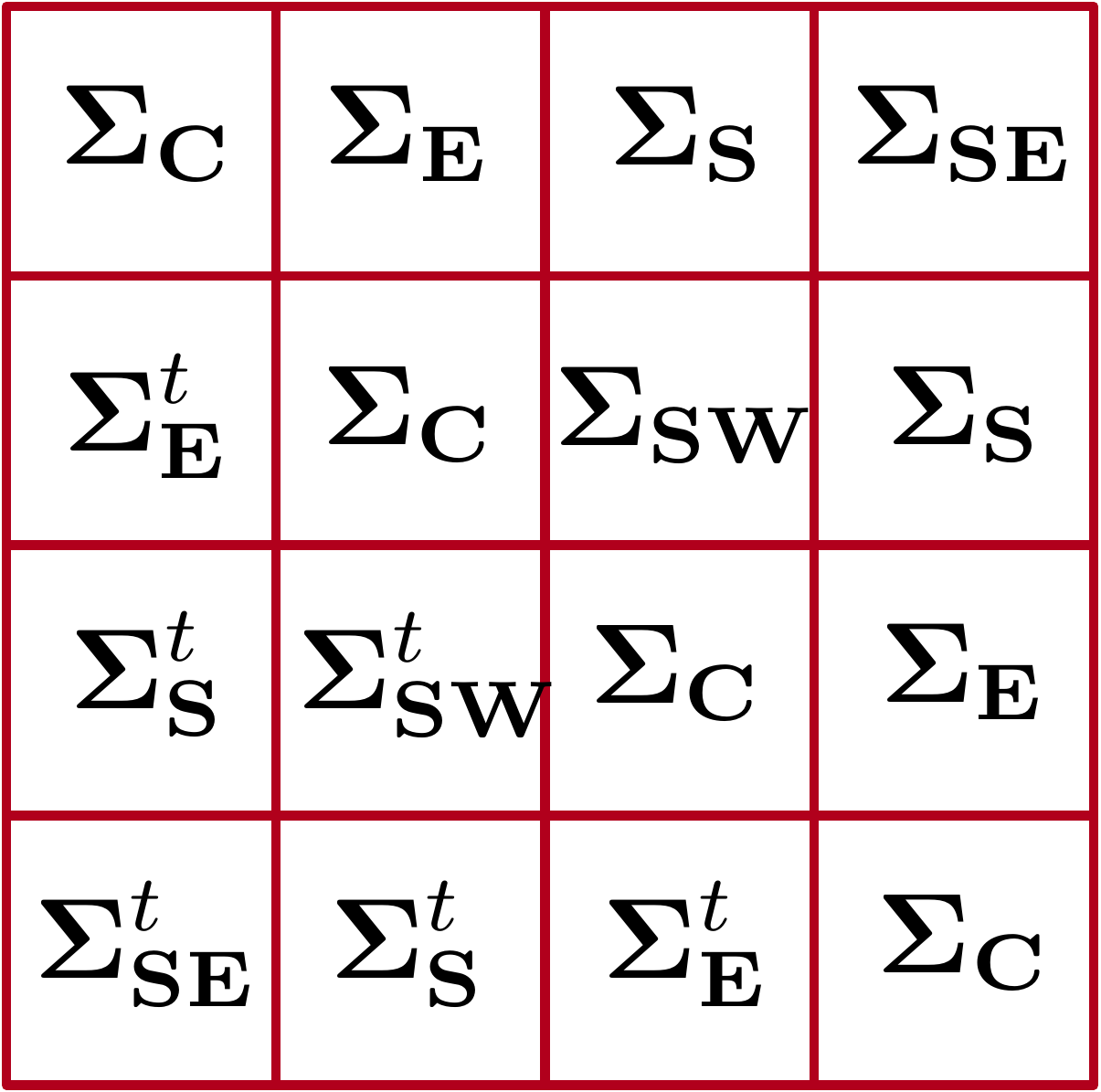}
\par\end{centering}
\label{fig:explained}}
\par\end{center}%
\end{minipage}
\par\end{centering}
\caption{(a) $256\times256$ covariance matrix of DCT coefficients of a color
sensor with bilinear demosaicking for an i.i.d signal (the correlation
values are thresholded for visualization purposes). (b): scan order
by blocks and coefficients. (c): types of sub-matrices representing
the 9 covariance matrices.\label{fig:(a)-Covariance-matrix}}
\end{figure}

As non connected blocks are uncorrelated, we focus here on only four
adjacent $8\times8$ blocks of unquantized DCT coefficients, as depicted
in Figure~(\ref{fig:scan}). This selection enables us to analyze
correlations within a block, but also correlations between horizontal,
vertical and diagonal neighboring blocks. By observing Figure~(\ref{fig:color})
together with the scan order depicted in Figure~(\ref{fig:scan}),
we can decompose the entire covariance matrix into four types of matrices
of size $64\times64$ as illustrated in Figure~(\ref{fig:explained}):
\begin{itemize}
\item Intra-block $8\times8$ covariance matrices of type $\boldsymbol{\Sigma}_{C}$
capture the correlations between DCT coefficients in the same block.
They are located on the diagonal of the covariance matrix $\boldsymbol{\Sigma}_d$.
Note that DCT coefficients can be positively or negatively correlated.
\item Horizontal inter-block covariance matrices of type $\boldsymbol{\Sigma}_{E}$
or $\boldsymbol{\Sigma}_{W}$. They hold correlations between horizontal
blocks.
\item Vertical inter-block covariance matrices capture correlations between
vertical blocks. They can be of type $\boldsymbol{\Sigma}_{N}$ or
$\boldsymbol{\Sigma}_{S}$.
\item Diagonal inter-block covariance matrices capture correlations between
diagonal blocks. They can be of type $\boldsymbol{\Sigma}_{NE}$,
$\boldsymbol{\Sigma}_{SW}$,$\boldsymbol{\Sigma}_{SE}$, or $\boldsymbol{\Sigma}_{NW}$.
\end{itemize}
It is worth noting that the stationary behavior that appears here
in $\mathbf{\Sigma}_d$ is not true for real images where the input
signal is not identically distributed. Being aware of this, we do
not consider stationarity for the embedding procedure (see Section~(\ref{sec:Embedding-Scheme})) but we use it only for analysis purposes. We give now an accurate analysis of the structure of the above defined covariances matrices.

\subsection{Intra-block correlations}

The coefficients of the covariance matrix for intra-block correlations
are of two types: they are either due to demosaicking artifacts (see
Section~\ref{subsec:Effect-of-demosaicking}), or the consequence
of low-pass filtering (see Section~\ref{subsec:Effect-of-LP}).

\subsubsection{Effect of demosaicking\label{subsec:Effect-of-demosaicking}}

\begin{figure}[h]
\begin{centering}
\subfloat[\label{fig:dem:a}]{\begin{centering}
\includegraphics[width=0.45\columnwidth]{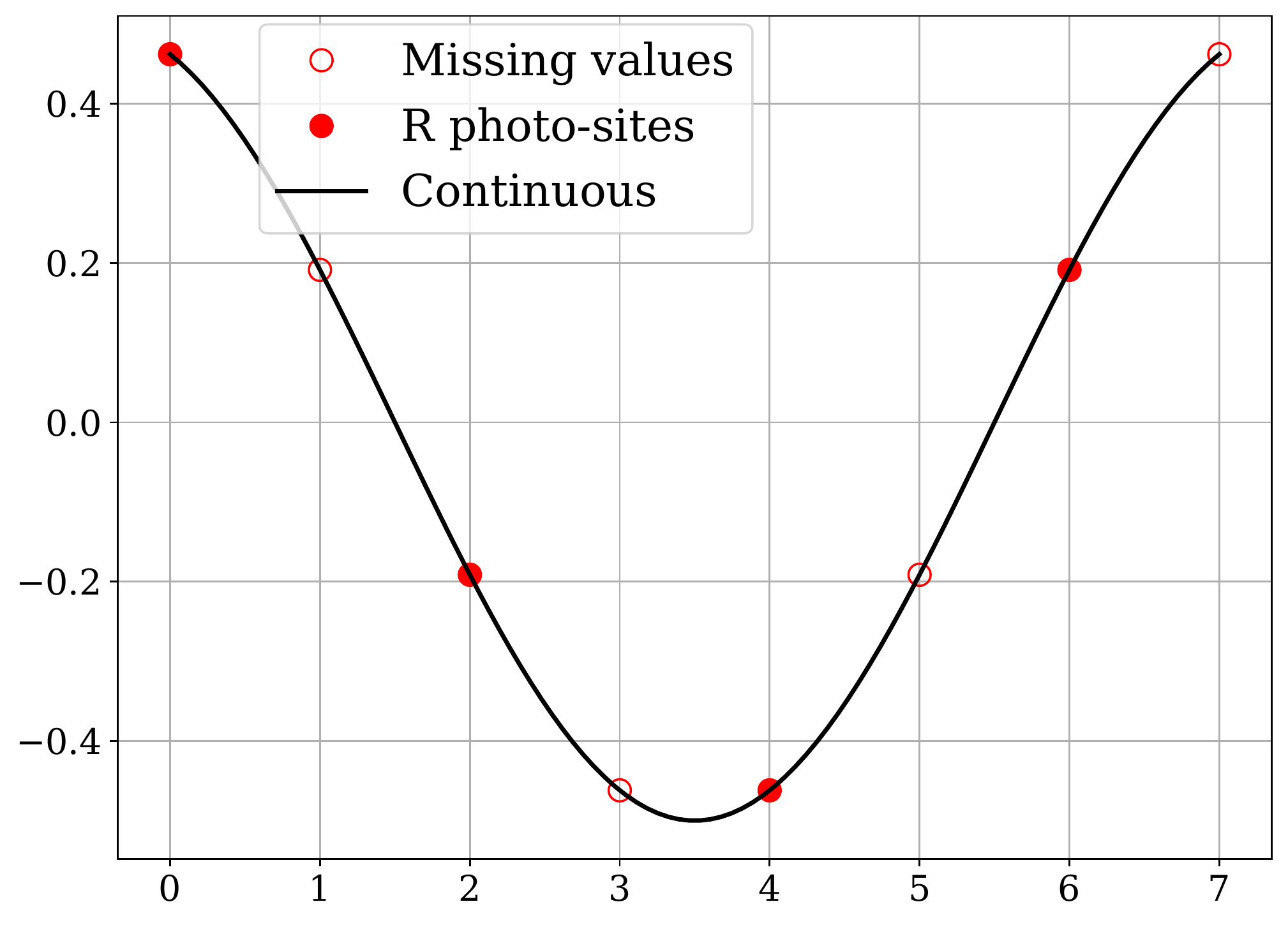}
\par\end{centering}
}\subfloat[\label{fig:dem:b}]{\begin{centering}
\includegraphics[width=0.45\columnwidth]{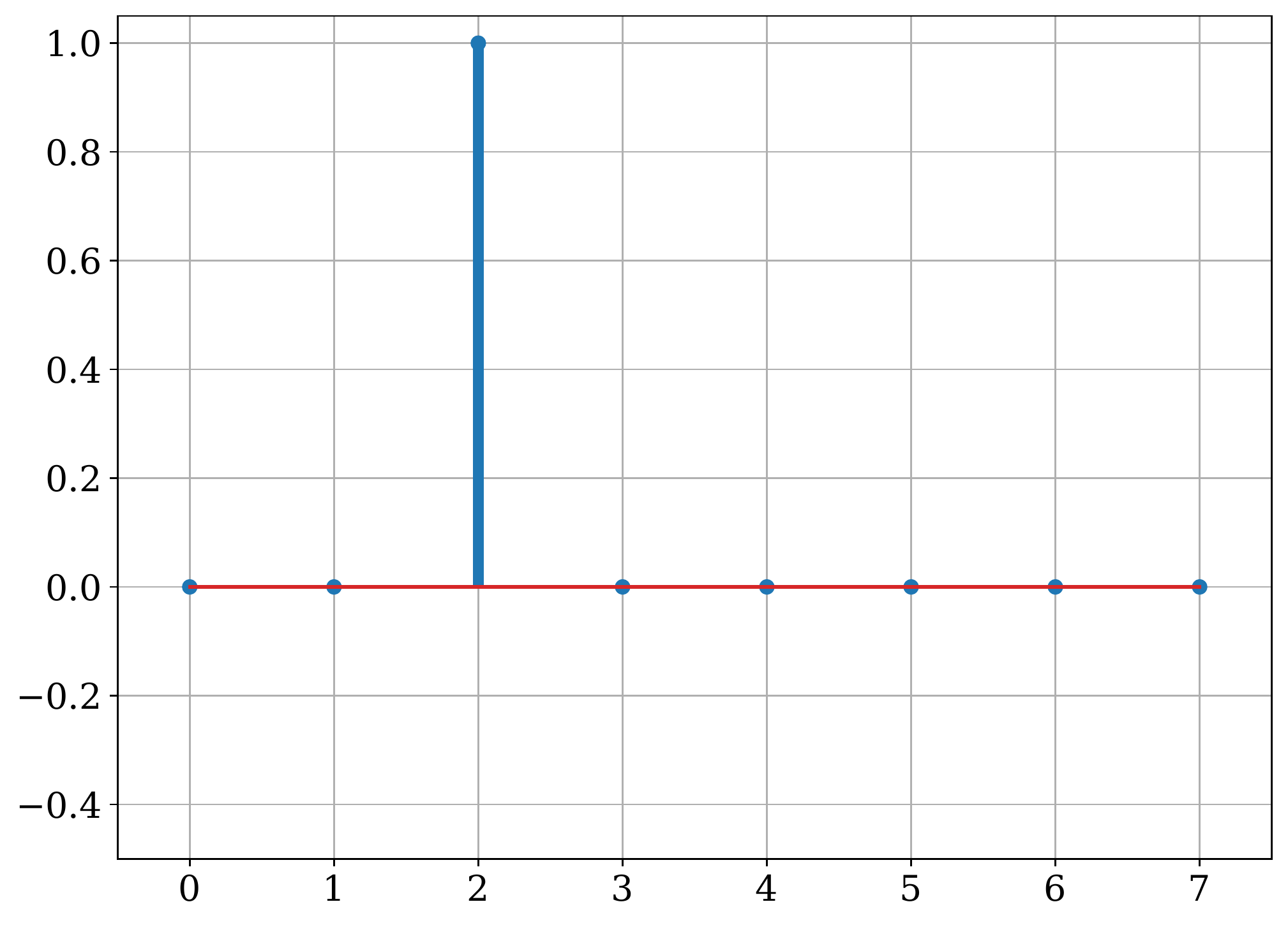}
\par\end{centering}
}
\par\end{centering}
\subfloat[\label{fig:dem:c}]{\begin{centering}
\includegraphics[width=0.45\columnwidth]{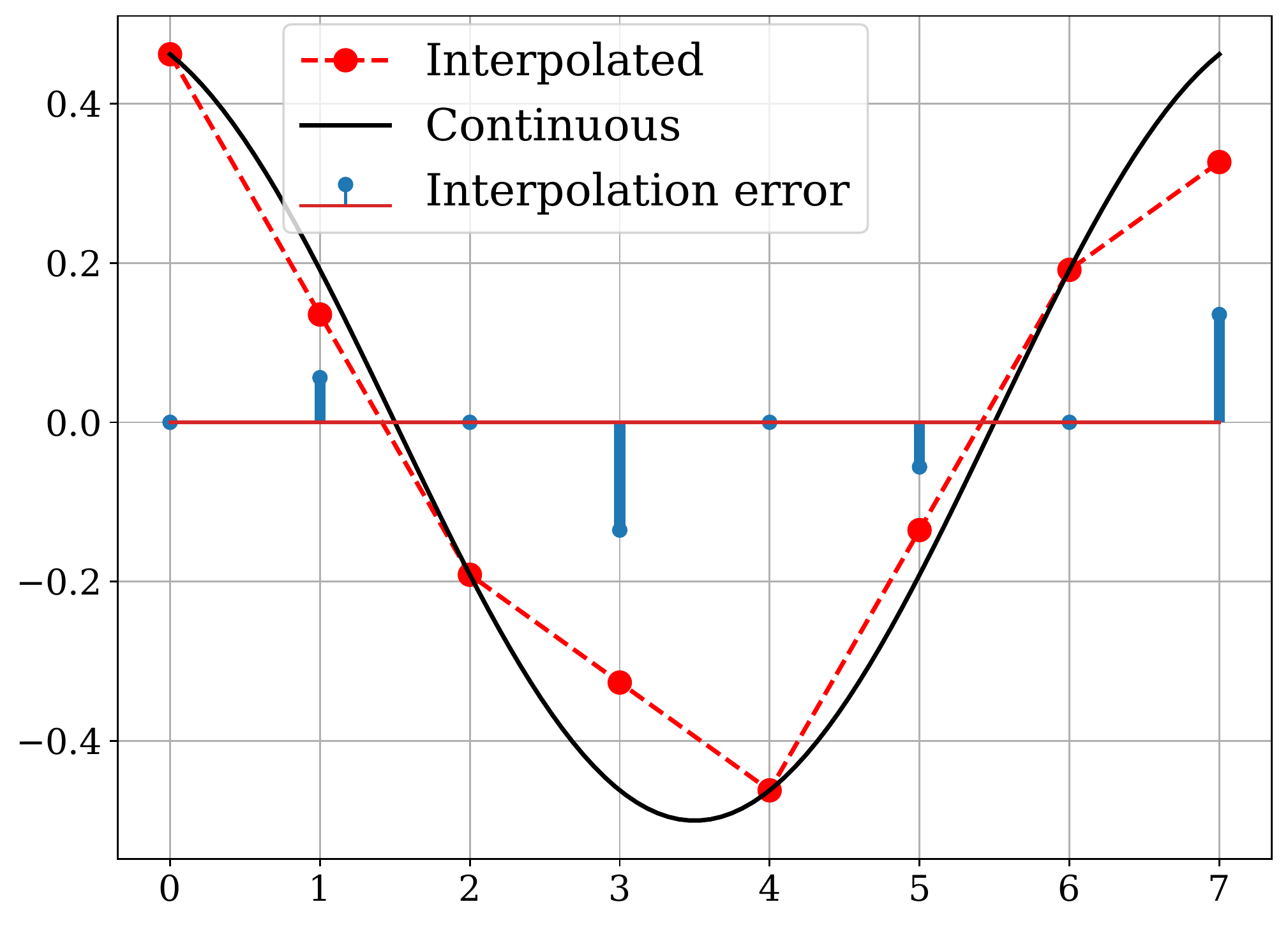}
\par\end{centering}
}
\begin{centering}
\subfloat[\label{fig:dem:d}]{\begin{centering}
\includegraphics[width=0.95\columnwidth]{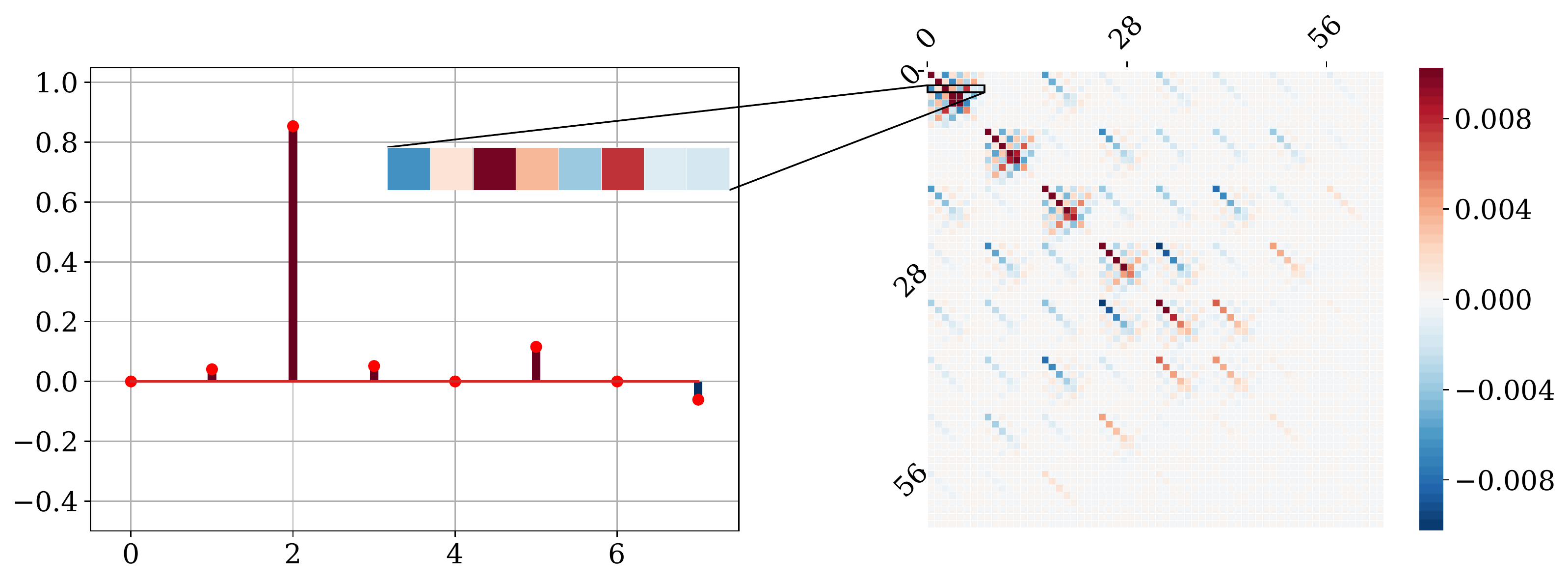}
\par\end{centering}
}
\par\end{centering}
\caption{Impact of demosaicking on correlation between intra-block DCT coefficients:
(a) visualization of one line $\mathbf{b}_{c}$ of the $(0,2)$ mode
in the spatial domain. (b) $DCT(\mathbf{b}_{c})$ . (c) Continuous
signal, interpolated signal $\mathbf{b}_{i}$ and interpolation error.
(d) comparison between the DCT transform of the interpolated waveform
(left) and the covariance matrix obtained from interpolated pure DCT
modes (right).\label{fig:(a)-INTRA-1D-interpolation-errors-4fig}}
\end{figure}

In order to emphasize the effect of demosaicking, we select only one color channel, the red one, and we investigate the intra-block correlations when the luminance computation operation is not taken into account.
The demosaicking operation introduces dependencies within the same block and this is both due to the structure of the CFA itself and the color interpolation algorithm. For a given waveform of the DCT mode , i.e. its representation in the spatial domain\footnote{a.k.a. the pixel domain.}, the demosaicking operation, which can be seen as a succession of sub-sampling and linear interpolation, introduces artifacts coming from interpolation errors, such that the final result is a linear combination of the other 63 DCT modes.
The initial mode is encoded with a larger magnitude than the others as summed up in the following expression:

\[
\mathrm{DCT}\left(\mathrm{Dem}\left(\mathrm{mode_{i}}\right)\right)=A_{i}\cdot\mathrm{mode_{i}}+\underbrace{\sum_{i\neq j}A_{j}\cdot\mathrm{mode_{j}}}_{\mathrm{DCT\:artifacts}},
\]
here $\mathrm{mode}_{i}$ represents the spatial representation of
DCT mode $i$ after demosaicking (the $\mathrm{Dem()}$ function).
The appearance of the $A_{j}$ terms is due to small interpolation
errors of mode $i$. These artifacts are illustrated in Figure~\ref{fig:(a)-INTRA-1D-interpolation-errors-4fig}.
This figure can be explained as follows: in order to encode continuous waveforms that are interpolated during the demosaicking process, the interpolation process has to deal with missing values (see Figure \ref{fig:dem:a}), which encode other frequencies in the DCT domain (see Figure~\ref{fig:dem:c}). So, instead of encoding one component (see Figure~\ref{fig:dem:b}), it also encodes other DCT components (see Figure~\ref{fig:dem:d}). 

In Figure~\ref{fig:dem:d}, we also compare the covariance matrix computed by interpolating only the red channel on continuous DCT waveforms and the DCT of the interpolated waveform. Note that the fourth line of the covariance matrix is very similar with the components depicted in Figure~\ref{fig:dem:d}.

In the 2D spatial domain, for a single mode applied to a $8\times8$
photo-sites array, the demosaicking algorithm creates artifacts such
that the resulting image in the DCT domain is a linear mixture of
the different DCT modes.

\subsubsection{Effect of low pass filtering\label{subsec:Effect-of-LP}}

The second category of artifacts is due to a low-pass filter, which can be related to the conversion from RGB to luminance or to any downsampling operation. In order to simulate the effect of low pass filtering, we use a random independent noise as a RAW image and convolve this input
with a standard low pass filter, such as:
\[
L=\frac{1}{12}\cdot\left[\begin{array}{ccc}
1 & 1 & 1\\
1 & 4 & 1\\
1 & 1 & 1
\end{array}\right].
\]

\begin{figure*}[t]
\begin{centering}
\subfloat[\label{subfig:demo}]{\begin{centering}
\includegraphics[width=0.25\textwidth]{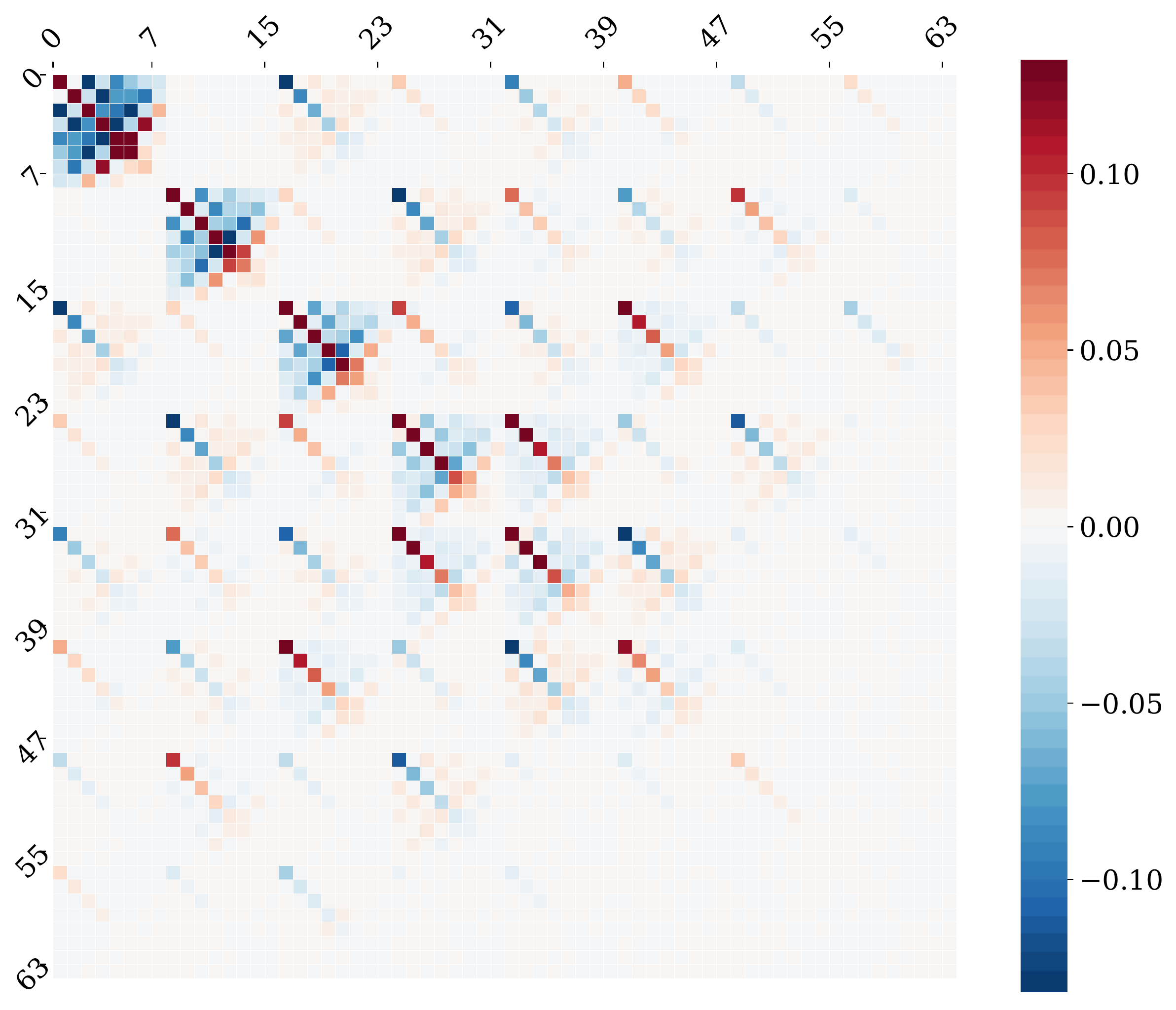}
\par\end{centering}
}\subfloat[\label{subfig:LP}]{\begin{centering}
\includegraphics[width=0.25\textwidth]{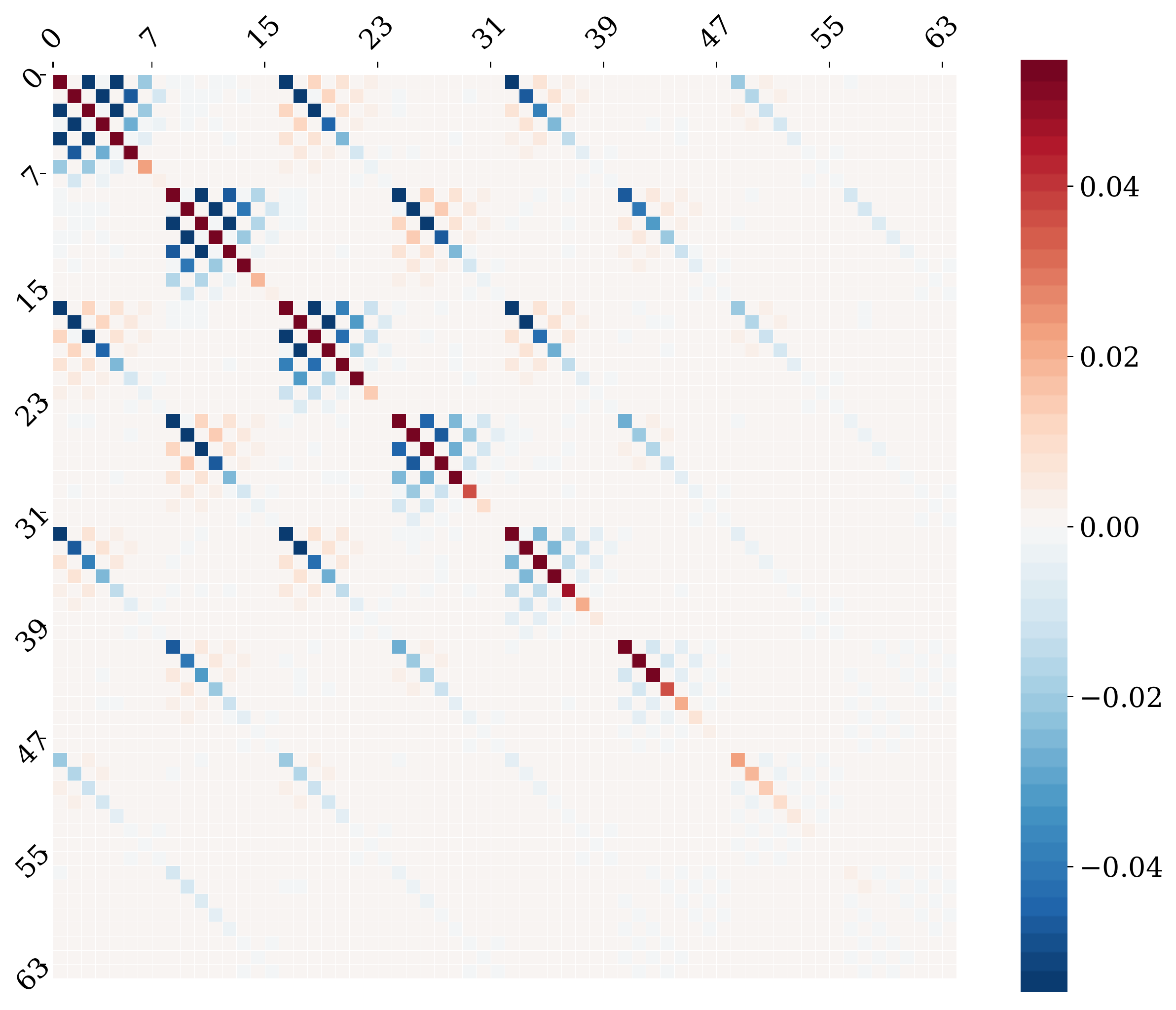}
\par\end{centering}
}\subfloat[\label{subfig:Tot}]{\begin{centering}
\includegraphics[width=0.25\textwidth]{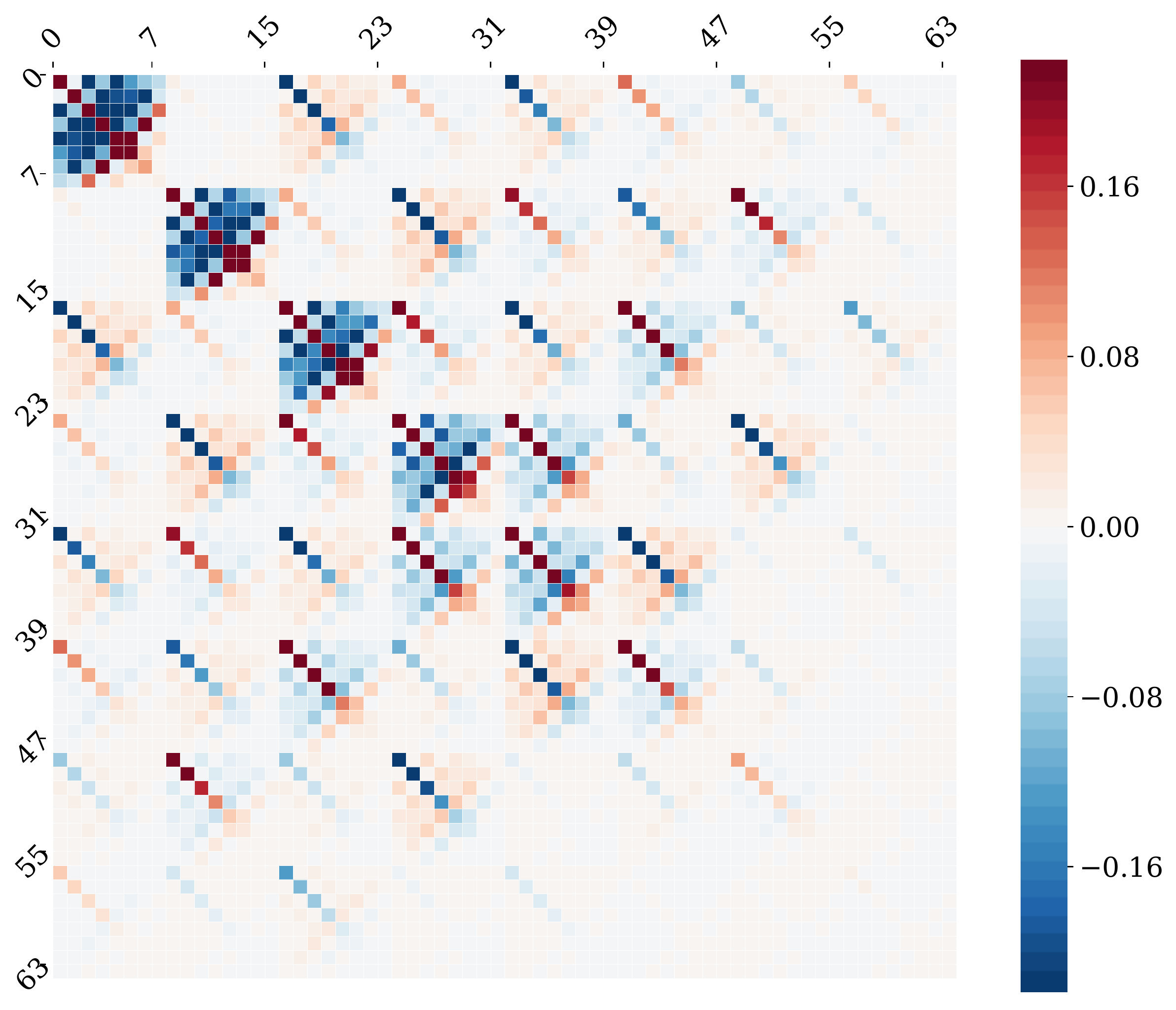}
\par\end{centering}
}
\par\end{centering}
\caption{(a): Covariance matrix computed after randomly generating DCT continuous
modes that are interpolated using bilinear filtering. (b): Intra correlations
within a block after low-pass filtering using filter $L$. (c): Intra-block
covariance matrix for $\mu=\mathrm{const}$.\label{fig:IntraCov}The
correlation values are thresholded for visualization purposes.}
\end{figure*}

The covariance matrix obtained by incorporating the low-pass filter in the development process is complementary to the covariance matrix obtained considering only the demosaicking artifacts. Figure~\ref{fig:IntraCov} shows these relationships: the total intra-covariance matrix (Figure~\ref{subfig:Tot}) can be approximated as the superposition of the covariance matrix of signals representing the demosaicking artifacts (Figure~\ref{subfig:demo}) and the covariance matrix of the independent signal at the photo-site level undergoing low-pass filtering (Figure~\ref{subfig:LP}).

\subsection{Inter-block correlations\label{subsec:Inter-block}}

\begin{figure}[H]
\subfloat[$(1,0)$]{\begin{centering}
\includegraphics[width=0.2\columnwidth]{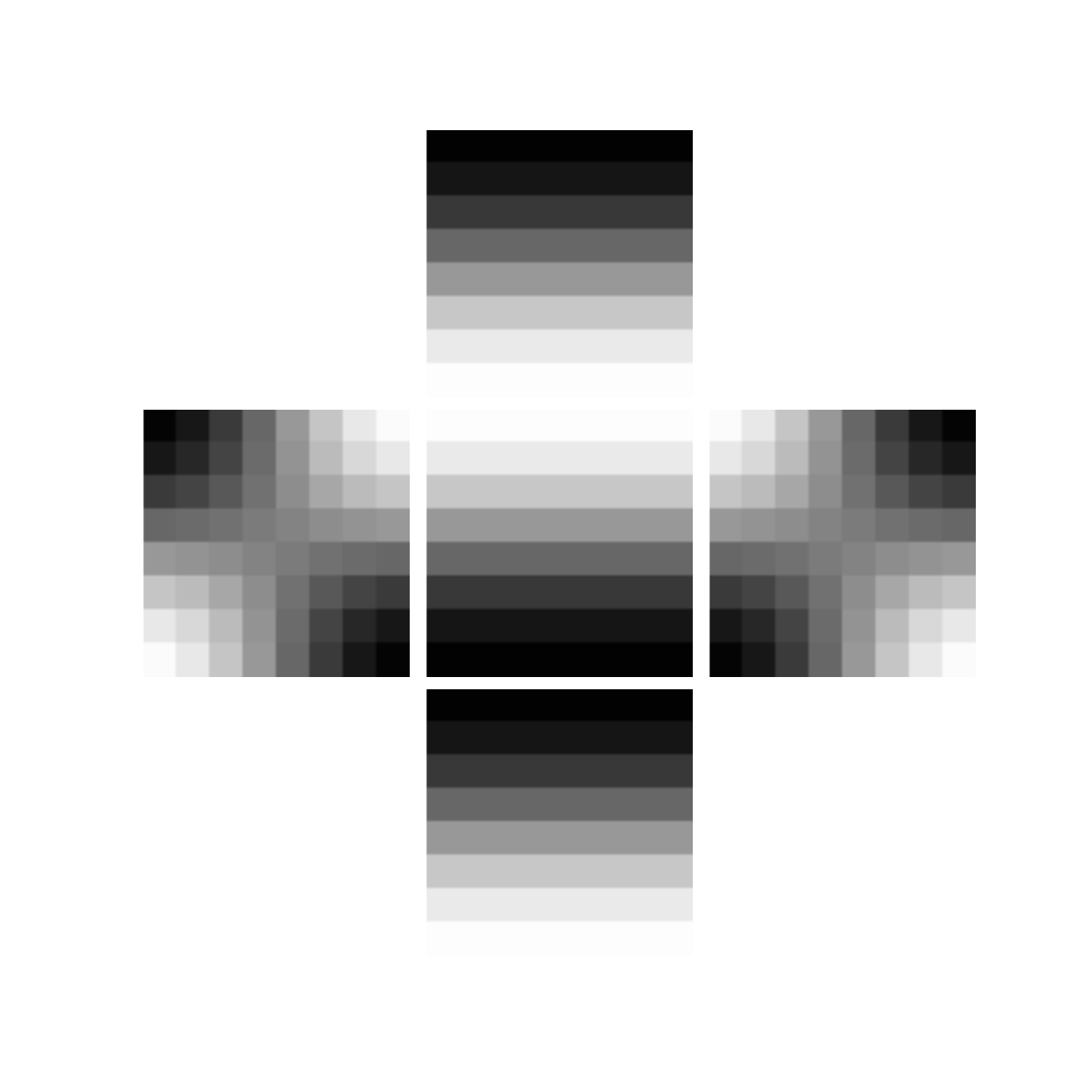}
\par\end{centering}
}\subfloat[$(0,1)$]{\begin{centering}
\includegraphics[width=0.2\columnwidth]{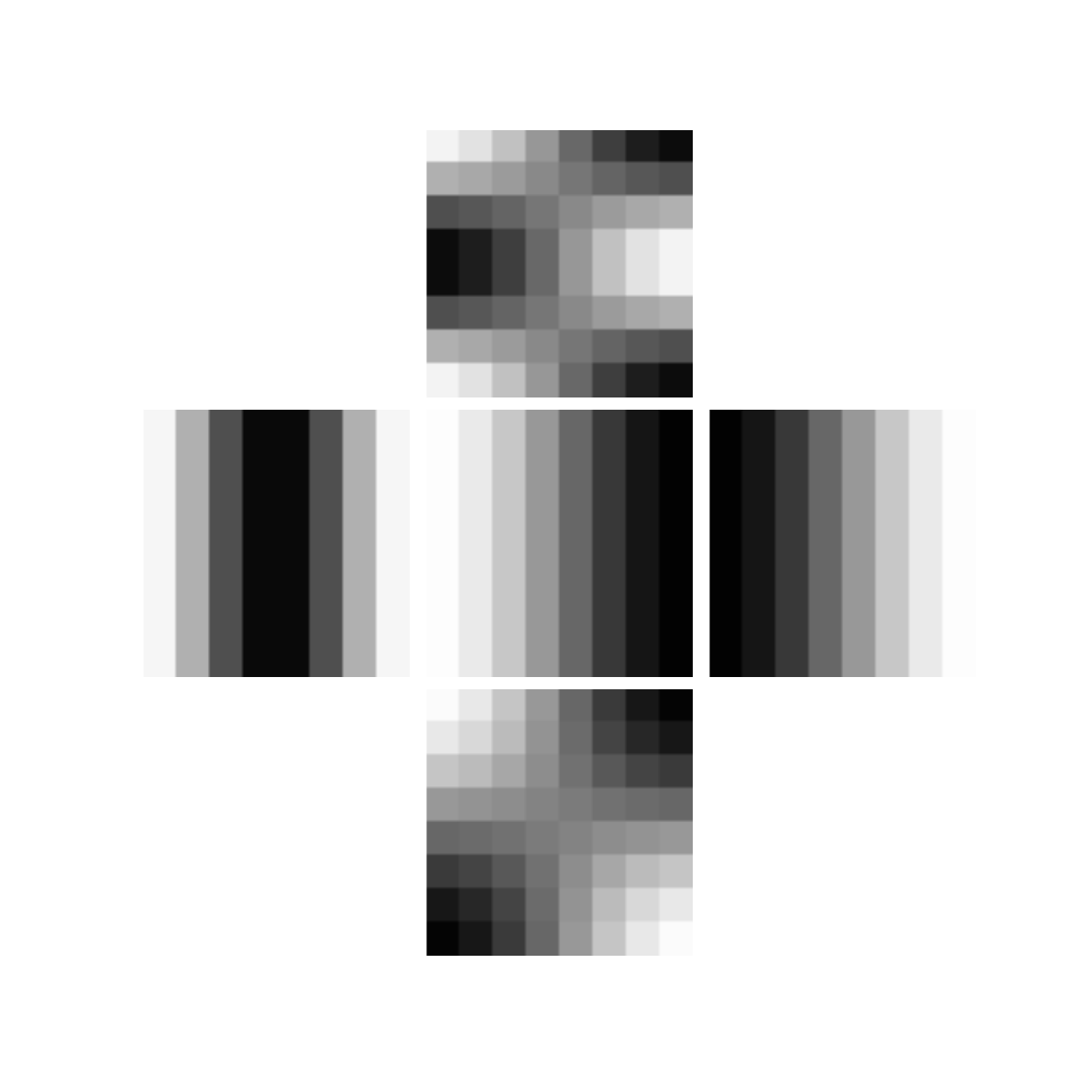}
\par\end{centering}
}\subfloat[$(0,5)$]{\begin{centering}
\includegraphics[width=0.2\columnwidth]{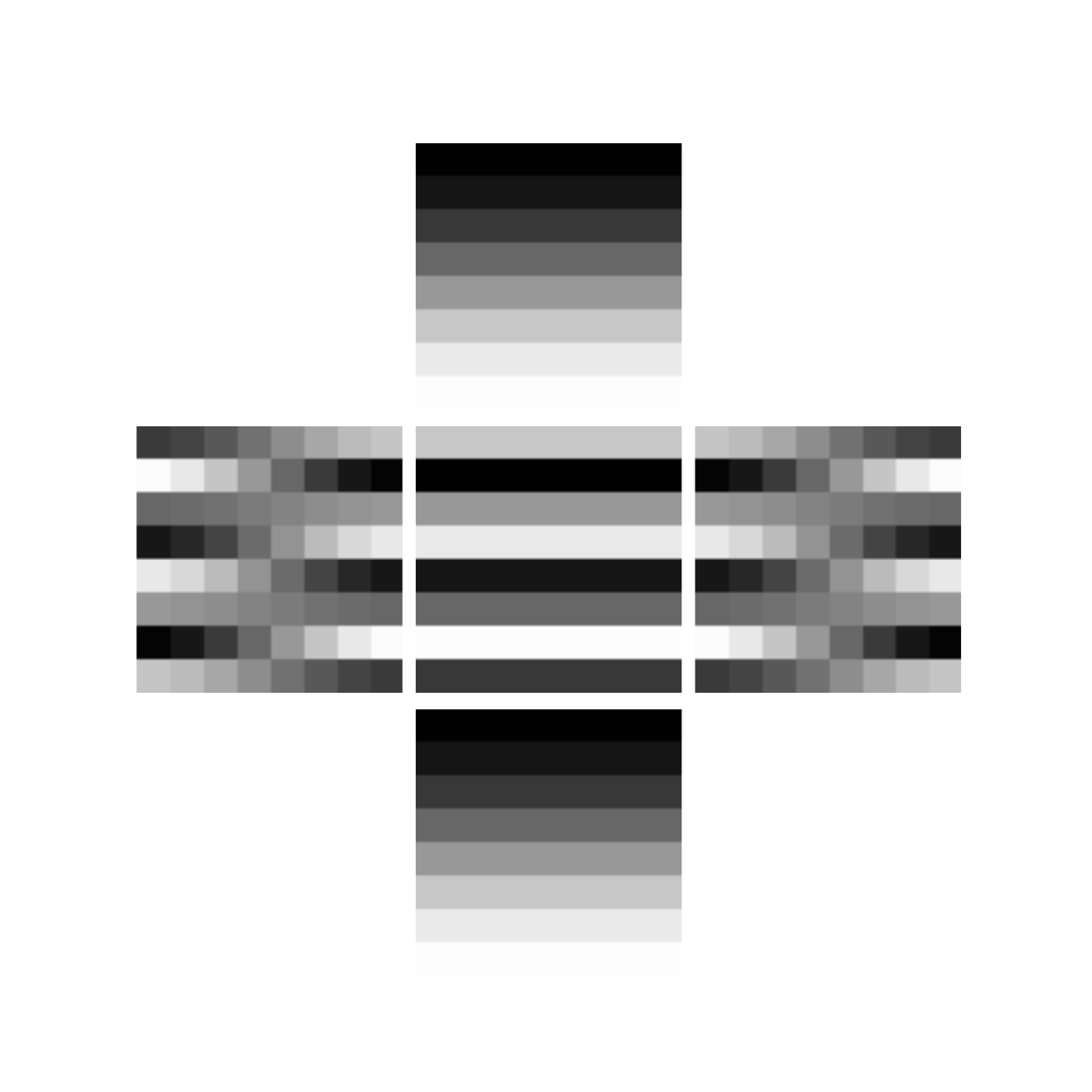}
\par\end{centering}
}\subfloat[$(1,1)$]{\begin{centering}
\includegraphics[width=0.2\columnwidth]{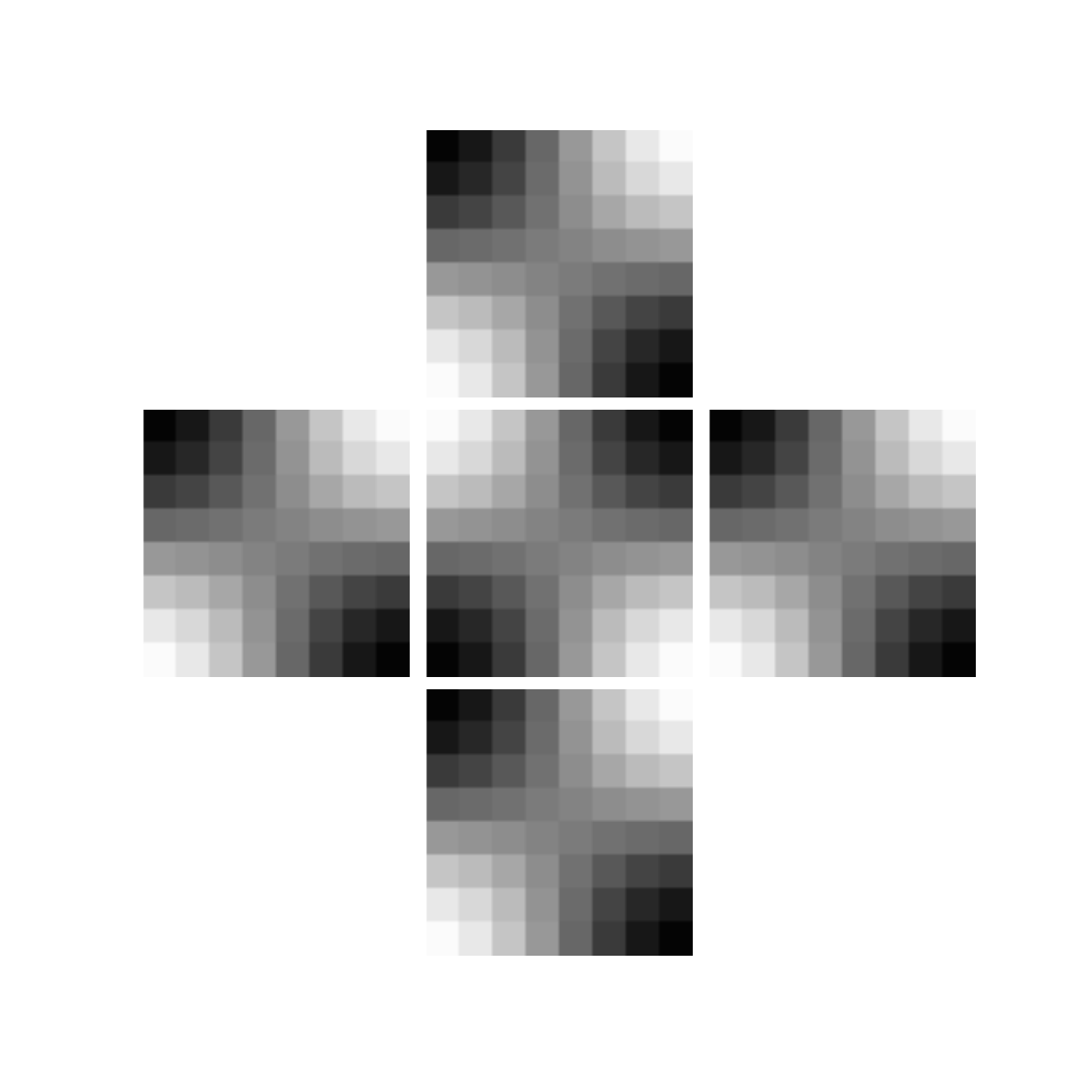}
\par\end{centering}
}
\centering{}\caption{Different DCT modes (central blocks) and their most correlated modes (represented by horizontal or vertical blocks). The presented locations of the blocks correspond to the spatial locations of the blocks. We can notice that the most correlated blocks preserve continuities between neighboring blocks.\label{fig:Different-arrangements-of}}
\end{figure}

Inter-block correlations between DCT coefficients are also caused
by demosaicking, which averages adjacent photo-site values to interpolate
the missing color values. It creates correlations between neighboring
pixels, including pixels belonging to two different DCT blocks. This
interpolation process highlights the low-pass component of the sensor
noise, and this is consistent across different demosaicking methods
(see~\cite{taburet:2019}). This phenomenon is illustrated in Figure~\ref{fig:Different-arrangements-of},
which shows for different DCT modes in the spatial domain, the arrangements
of blocks that are the most correlated for the horizontal and vertical
neighbors. For each arrangement, we can notice that the continuity
from one block to its neighbor is preserved.

The most significant correlations correspond to the surrounding vertical
and horizontal blocks. This is due to the large number of neighboring
photo-sites involved in the interpolation process. Note that the largest
correlations are for the same vertical or horizontal frequency due
to frequencies consistency between adjacent blocks.

The sign of the correlations represents the preservation of continuity
between blocks in order to guarantee spatial continuity. For example,
alternating signs are due to the topology of the waveforms. For example for mode $(1,0)$, all modes $(i,0)$ have a white top line but the bottom line alternates between white and black w.r.t. $i$.

It is interesting to connect this analysis with the recent steganographic
scheme proposed by Li \emph{et al.}~\cite{li2018defining} which
synchronizes embedding changes between several DCT modes by empirically
adjusting costs in order to favor continuities between blocks. This
practical rationale is now theoretically justified by our analysis.

\section{Simulated Embedding\label{sec:Embedding-Scheme}}

In order to perform simulated embedding, we first need to compute
the probability mass function (pmf) of the embedding changes for each
coefficient of the cover JPEG before performing embedding changes.
We then sample according to this pmf in order to generate the quantized
stego signal $\mathbf{\tilde{s}}_{d}$ and consequently the JPEG stego
image. We recall that true embedding may also be performed by computing
the costs associated with each embedding probability change, and by
running a multilayer STC (see Section~\ref{subsec:Pseudo-embedding,-simulated-embe}).

In Section~(\ref{sec:Modeling-dependencies-in}), we saw that in
the DCT domain, the of the coefficients resulting from a stego
signal follows a zero-mean multivariate Gaussian distribution. Its covariance matrix
computed for $3\times3$ blocks (each block containing $8\times8$
DCT coefficients) is given by~(\ref{eq:Cov_from_RAW}). Moreover
8-connected blocks are correlated, but two not connected blocks can
be drawn independently. 

In order to sample according to the joint
distribution, we need to compute conditional pmfs for each quantized
DCT coefficient using the four following technical developments:
\begin{enumerate}
\item The decomposition of the image in the DCT domain into four disjoint
macro lattices (see~(\ref{subsec:Decomposition-into-lattices})).
\item The use of the chain rule of conditional sampling (see~(\ref{subsec:Conditional-sampling})) combined with an embedding over $4 \times 64$ lattices.
\item The computation of the associated probability mass functions and associated sampling operations in the continuous and quantized domain (see~\ref{subsec:pmf}).
\item The computation of the embedding capacity (see~\ref{subsec:ent}).
\end{enumerate}

\subsection{Decomposition into lattices\label{subsec:Decomposition-into-lattices}}

The embedding has to take into account three facts:
\begin{enumerate}
\item Intra-block dependencies within each $8\times8$ block.
\item Inter-block dependencies between one central block and its horizontal,
vertical and diagonal neighbors.
\item Independence of blocks that are not neighbors.
\end{enumerate}
Argument (1) means that we practically have to use 64 lattices (one
per DCT mode) to perform embedding in one DCT block and (2) and (3)
mean that we need a maximum of four macro-lattices $\{\Lambda_{1},\Lambda_{2},\Lambda_{3},\Lambda_{4}\}$
to perform embedding in each DCT block while respecting the correlations
exhibited by the computed covariance matrix. 

\begin{figure}[h]
\begin{centering}
\includegraphics[width=0.6\columnwidth]{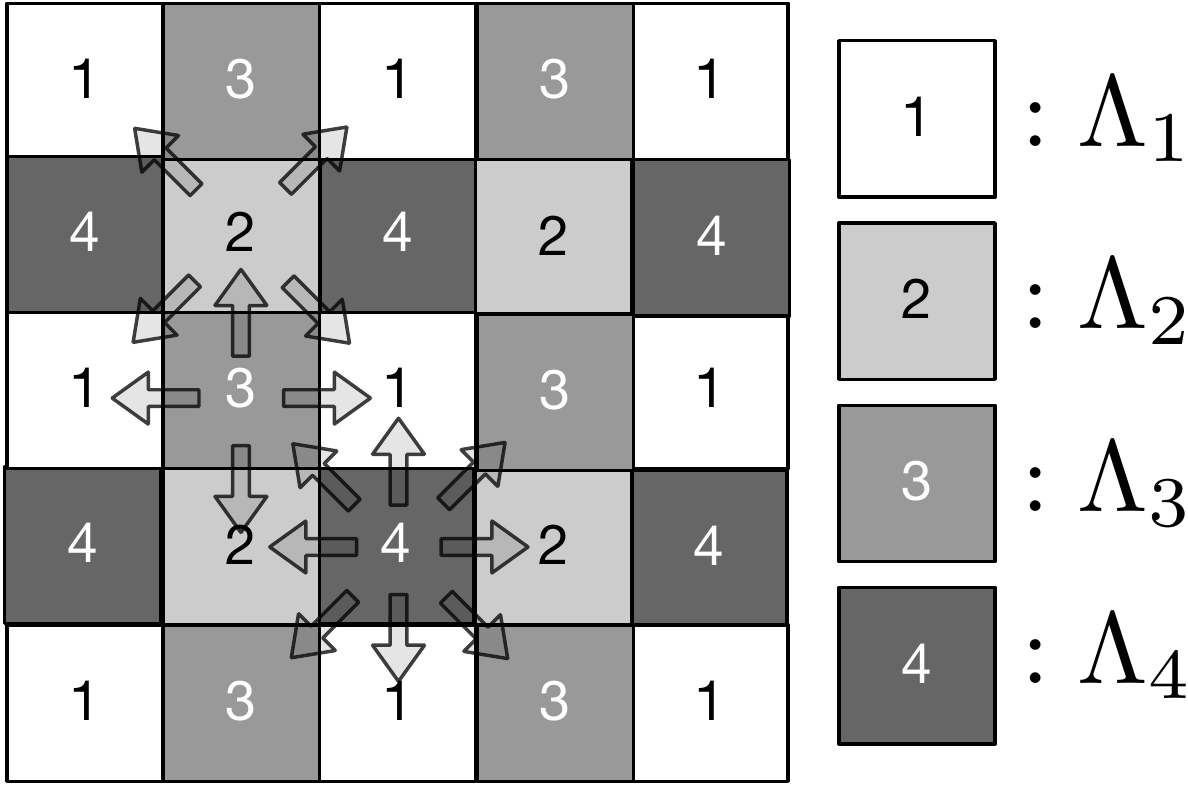}
\par\end{centering}
\centering{}\caption{The four macro lattices used for embedding. Arrows indicate the neighborhood
used to compute conditional probabilities.\label{fig:Lattices}}
\end{figure}

The different macro-lattices
are illustrated in Figure~\ref{fig:Lattices} together with the neighboring
blocks that are involved.

Condider a vector of $3\times 3$ blocks of the stego signal in the DCT domain. Let $\mathbf{s}_d^{C}$ be the central block and $\mathbf{s}_d^{NW},\mathbf{s}_d^{N},\mathbf{s}_d^{NE},\mathbf{s}_d^{W},\mathbf{s}_d^{E},\mathbf{s}_d^{SW},\mathbf{s}_d^{S}$, $\mathbf{s}_d^{SE}$ be respectively the north-west, north,
north-east, west, east, south-west, south, and south-east blocks w.r.t. the central one. 

We can build the vector of interest $\mathbf{s}^{\star}$, used to compute conditional probabilities (see next sub-section), as follows:

- For $\Lambda_{1}$, only the intra-block covariance matrix is necessary,
computed w.r.t. $\mathbf{s}^{\star}=\mathbf{s}_d^{C}$,

- For $\Lambda_{2}$, $\mathbf{s}^{\star}=[\mathbf{s}_d^{C},\mathbf{s}_d^{NW},\mathbf{s}_d^{NE},\mathbf{s}_d^{SW},\mathbf{s}_d^{SE}]$,

- For $\Lambda_{3}$, $\mathbf{s}^{\star}=[\mathbf{s}_d^{C},\mathbf{s}_d^{N},\mathbf{s}_d^{W},\mathbf{s}_d^{E},\mathbf{s}_d^{S}]$,

- For $\Lambda_{4}$, $\mathbf{s}^{\star}=[\mathbf{s}_d^{C},\mathbf{s}_d^{NW},\mathbf{s}_d^{N},\mathbf{s}_d^{NE},\mathbf{s}_d^{W},\mathbf{s}_d^{E},\mathbf{s}_d^{SW},\mathbf{s}_d^{S},\mathbf{s}_d^{SE}]$.

We end up with a decomposition of the image into $4\times64=256$
lattices (four macro lattices and one lattice per DCT mode). In each
lattice, the covariance matrix may be expressed as:

\begin{equation}
\boldsymbol{\Sigma}_d=
	\begin{bmatrix*}[l]
		\boldsymbol{\Sigma}_{[0:64][0:64]} & \boldsymbol{\Sigma}_{[0:64][64:n\times64]}\\
		\boldsymbol{\Sigma}_{[64:n\times64][0:64]} & \boldsymbol{\Sigma}_{[64:n\times64][64:n\times64]}
	\end{bmatrix*},\label{eq:Shurr}
\end{equation}

with $n$ denoting the number of blocks in $\mathbf{s}^{\star}$ (see footnote \footnote{The pythonic notation $[i:j]$ means that all indexes from the interval$[i,j-1]$
are considered.}) and $n=1$ for $\Lambda_{1}$, $n=5$ for $\Lambda_{2}$ and $\Lambda_{3}$
and $n=9$ for $\Lambda_{4}$, see Figure~(\ref{fig:Lattices}).

\subsection{Conditional sampling\label{subsec:Conditional-sampling}}

Using the lattice decomposition, changes can be drawn independently
according to the pmf $\pi_{i}$ for simulated embedding in each lattice,
or using a STC based on costs $\rho_{i}$ (see (\ref{subsec:Pseudo-embedding,-simulated-embe})).
In order to derive the pmf $\pi_{i}(k)$ for each sample $i$ and
the modification magnitude $k$, we need to use conditional sampling,
a variation of Gibbs sampling, which enables to sample from a multivariate
distribution using only conditional distributions.

Without loss of generality, if we focus on the set of 4 macro lattices defined
in~(\ref{subsec:Decomposition-into-lattices}) (but this can be applied
on any number of lattices that are conditionally independent), the
chain rule of conditional probabilities gives 

\begin{equation*}\label{}
	\resizebox{\columnwidth}{!}{$
	\left.
	\begin{aligned}
		P\left(\mathbf{s}_d\right)&=P\left(\mathbf{\mathbf{s}}_{\Lambda_{1}},\mathbf{\mathbf{s}}_{\Lambda_{2}},\mathbf{s}_{\Lambda_{3}},\mathbf{\mathbf{s}}_{\Lambda_{4}}\right),\\
		&=P\left(\mathbf{s}_{\Lambda_{1}}\right)P\left(\mathbf{s}_{\Lambda_{2}}|\mathbf{\mathbf{s}}_{\Lambda_{1}}\right)P\left(\mathbf{s}_{\Lambda_{3}}|\mathbf{s}_{\Lambda_{1}},\mathbf{s}_{\Lambda_{2}}\right)P\left(\mathbf{s}_{\Lambda_{4}}|\mathbf{s}_{\Lambda_{1}},\mathbf{s}_{\Lambda_{2}},\mathbf{s}_{\Lambda_{3}}\right).
	\end{aligned}
	\right.
	$}
\end{equation*}
%$P\left(\mathbf{s}\right)=P\left(\mathbf{\mathbf{s}}_{\Lambda_{1}},\mathbf{\mathbf{s}}_{\Lambda_{2}},\mathbf{s}_{\Lambda_{3}},\mathbf{\mathbf{s}}_{\Lambda_{4}}\right)=P\left(\mathbf{s}_{\Lambda_{1}}\right)P\left(\mathbf{s}_{\Lambda_{2}}|\mathbf{\mathbf{s}}_{\Lambda_{1}}\right)P\left(\mathbf{s}_{\Lambda_{3}}|\mathbf{s}_{\Lambda_{1}},\mathbf{s}_{\Lambda_{2}}\right)P\left(\mathbf{s}_{\Lambda_{4}}|\mathbf{s}_{\Lambda_{1}},\mathbf{s}_{\Lambda_{2}},\mathbf{s}_{\Lambda_{3}}\right),$
where $\mathbf{s}$ is a random vector representing the whole set
of DCT coefficients related to the stego signal in the DCT domain, and $\mathbf{s}_{\Lambda_{i}}$ represents
the DCT coefficients belonging to lattice $\Lambda_{i}$. % $i \in \{1,\dots,4\}$.

This means that we can perform (simulated) embedding first in lattice
$\Lambda_{1}$ by sampling according to $P\left(\mathbf{s}_{\Lambda_{1}}\right)$,
then embed in the second lattice by sampling according to $P\left(\mathbf{\mathbf{s}}_{\Lambda_{2}}|\mathbf{s}_{\Lambda_{1}}\right)$
and so on until embedding in lattice $\Lambda_{4}$ by sampling according
to $P\left(\mathbf{\mathbf{s}}_{\Lambda_{4}}|\mathbf{s}_{\Lambda_{1}},\mathbf{s}_{\Lambda_{2}},\mathbf{s}_{\Lambda_{3}}\right).$\\

{\it Conditional distribution in the continuous domain:}\\

We explain now how we can compute the conditional probability related to a particular DCT coefficient.\\
For each macro lattice $\Lambda_{k}$, $k\in{1,..,4}$ and block $\ell$,
the random vector of stego signal components conditioned by the previous embeddings follows a Multivariate Gaussian Distribution:
$\mathcal{N}(\boldsymbol{\mathbf{m}}_{k,\ell},\mathbf{\Sigma}_{k,\ell})$,
where $\boldsymbol{\mathbf{m}}_{k,\ell}$ and $\mathbf{\Sigma}_{k,\ell}$
can be computed using the Schur complement of the full covariance
matrix(\ref{eq:Shurr}~\cite{von1964mathematical}). For example, if we perform the embedding in block $\ell$ from lattice
$\Lambda_{4}$, the mean vector $\boldsymbol{\mathbf{m}}_{4,\ell}$
and the covariance matrix $\mathbf{\Sigma}_{4,\ell}$ are computed
conditionally to the embedding performed in $\{\Lambda_{1},\Lambda_{2},\Lambda_{3}\}$
(recall that the mean of $\mathbf{s}_d$ is $0$):

\begin{equation}
\boldsymbol{\mathbf{m}}_{4,\ell}=\boldsymbol{\Sigma}_{[0:64][64:n\times64]}\boldsymbol{\Sigma}_{[64:n\times64][64:n\times64]}^{-1}\mathbf{s}_{\Lambda_{1},\Lambda_{2},\Lambda_{3}},\label{eq:conditional_mean}
\end{equation}
and the Schur complement is given by:
% \begin{equation}
% 	\begin{array}{ccc}
% 	\mathbf{\Sigma}_{4,\ell} & = & \boldsymbol{\Sigma}_{[0:64][0:64]}\boldsymbol{\Sigma}_{[0:64][64:n\times64]}\\
% 	 &  & \boldsymbol{\Sigma}_{[64:n\times64][64:n\times64]}^{-1}\boldsymbol{\Sigma}_{[64:n\times64][0:64]},
% 	\end{array}\label{eq:conditional_cov}
% \end{equation}

\begin{equation}
	\resizebox{\columnwidth}{!}{$
		\mathbf{\Sigma}_{4,\ell} = \boldsymbol{\Sigma}_{[0:64][0:64]}\boldsymbol{\Sigma}_{[0:64][64:n\times64]}\boldsymbol{\Sigma}_{[64:n\times64][64:n\times64]}^{-1}\boldsymbol{\Sigma}_{[64:n\times64][0:64]}
	$}\label{eq:conditional_cov}
\end{equation}

for the stego-signal $\mathbf{s}_{\Lambda_{1},\Lambda_{2},\Lambda_{3}}$
defined by the surrounding blocks belonging to the three first lattices
(see Figure~\ref{fig:Lattices}).

At this stage of the study, it is possible to generate the $64$ stego
signal values $\mathbf{s}_{k,\ell}=(c_{0},\ldots,c_{63})_{k,\ell}^{t}$
in the DCT domain. %Even if embedding is realized by sampling quantized values of the stego signal in the JPEG domain, we will explain later the need for also generating $\mathbf{s}_{k,\ell}$. 

For each of the
64 lattices in each macro lattice, we sample by using the Cholesky
decomposition of the corresponding covariance matrix $\mathbf{\Sigma}_{k,\ell}$,
denoted $\mathbf{L}_{k,\ell}$, which is a lower triangular matrix
such that $\mathbf{\Sigma}_{k,\ell}=\mathbf{L}_{k,\ell}\cdot\mathbf{L}_{k,\ell}^{t}$.

Let $(N_{1},\,N_{2},\,\cdots,\,N_{63})\sim\mathcal{N}(\mathbf{0},\mathbf{I}_{64})$
a standard multivariate Gaussian distribution, and $\mathbf{n}=(n_{0},\ldots,n_{63})$
an outcome of it. Then $\mathbf{s}_{k,\ell}\sim\mathcal{N}(\boldsymbol{\mathbf{m}}_{k,\ell},\mathbf{\Sigma}_{k,\ell})$
can be sampled by computing $\mathbf{s}_{k,\ell}=\mathbf{\boldsymbol{m}}_{k,\ell}+\mathbf{L}_{k,\ell}\mathbf{n}$. 
More precisely, because we need to generate $\mathbf{s}_{k,\ell}$ iteratively, omitting here indexes $(k,\ell)$ for writing convenience, we have:

\begin{equation*}\label{}
	\left\{
	\begin{aligned}
		s_{0}&=m_{0}+L\left(0,0\right)\cdot n_{0}\\
		s_{1|0}&=\underbrace{m_{1}+L\left(1,0\right)\cdot n_{0}}_{m_{1|0}}+\underbrace{L\left(1,1\right)}_{\sigma_{1|0}^{2}}\cdot n_{1}\\
			\vdots
	\end{aligned},
	\right.		
\end{equation*}

and

\begin{equation}
S_{i|i-1\ldots,0}\sim\mathcal{N}\left(m'_{i},\sigma_{i}'^{2}\right)\,\,\,\,\,1\leq i\leq63,\label{eq:Cond_distrib}
\end{equation}
with $m'_{i}=m{}_{i}+\sum_{l=0}^{i-1}L(i,l)n_{l}$, and $\sigma_{i}'^{2}=L^{2}(i,i)$,
$i\geq1$,~$m'_{0}=m{}_{0}$, $\sigma'^{2}{}_{i}=L^{2}(0,0)$. 

Equation~(\ref{eq:Cond_distrib}) gives consequently the conditional distribution of each sample of the stego signal in the continuous
domain.

\subsection{Computation of the probability mass functions and sampling}\label{subsec:pmf}

Using the JPEG quantization matrix, the stego signal undergoes a quantization
and the conditioned probability density function has to be to converted
into a probability mass function which takes into account the associated
quantization table for the chosen quality factor $QF$. To compute
$\pi_{i}(k)=\mathrm{Pr}[\bar{S_{i}}=k]$, the probability that the
stego signal produces a change of magnitude $k\in\mathbb{Z}$ at a
coefficient $i\in\mathbb{N}$ for a given block, we compute the quantized
version of the real valued random variable $S_{i}$. This probability
mass function is given by:

\begin{align}
\pi_{i}(k) & =\mathrm{Pr}\Big[u_{k}<\frac{S_{i}}{Q_{i}}\leq u_{k+1}\Big],\nonumber \\
 & =\int_{u_{k}}^{u_{k+1}}{\frac{1}{\sqrt{2\pi\hat{\sigma}_{i}^{2}}}\mathrm{exp}\Big(-\frac{(x-\hat{m}_{i})^{2}}{2\hat{\sigma}_{i}^{2}}\Big)\mathrm{d}x},\nonumber \\
 & =\frac{1}{2}\Big[\mathrm{erf}\Big(\frac{u_{k+1}-\hat{m}_{i}}{\sqrt{2}\hat{\sigma}_{i}}\Big)-\mathrm{erf}\Big(\frac{u_{k}-\hat{m}_{i}}{\sqrt{2}\hat{\sigma}_{i}}\Big)\Big],\label{eq:ERF}
\end{align}
where $u_{k}=[\hat{m}_{i}]-0.5+k$, $\hat{m}_{i}=m'_{i}/Q_{i}$, $\hat{\sigma}_{i}=\sigma'_{i}/Q_{i}$
for parameters $m'_{i}$ and $\sigma'_{i}$ before quantization associated
with a quantization step $Q_{i}$. At each step $i$, the parameters
$m'_{i}$ and $\sigma'_{i}$ have to be generated in the continuous
domain with the knowledge of values drawn at steps $0\leq l\leq i-1$.
All the previous continuous samples are then needed to compute $m'_{i}$
and $\sigma'_{i}$. Once a sample has been generated in the discrete
domain, we need then to obtain a candidate in the continuous domain
which could have led to the sampled discrete value. This could be
done for example by using rejection sampling, where we can obtain
for each discrete sample its continuous candidate $S_{i\mid\bar{c}_{i}}$.

Rejection sampling works in the following way: for each discrete sampled value, we sample according to the continuous distribution until we find the appropriate candidate $S_{i}|\bar{s}_{i}$ such that:

\begin{equation}
u_{k}<S_{i}|\bar{s}_{i}<u_{k+1}.\label{eq:RejecSampling}
\end{equation}
where $\bar{s}_{i}=k$, $u_{k}=[\hat{m}_{i}]-0.5+k$, and $k\in \mathbb{Z}$
the symbol sampled as a modification in the discrete domain.

\begin{figure}[h!]
\begin{centering}
\includegraphics[width=0.8\columnwidth]{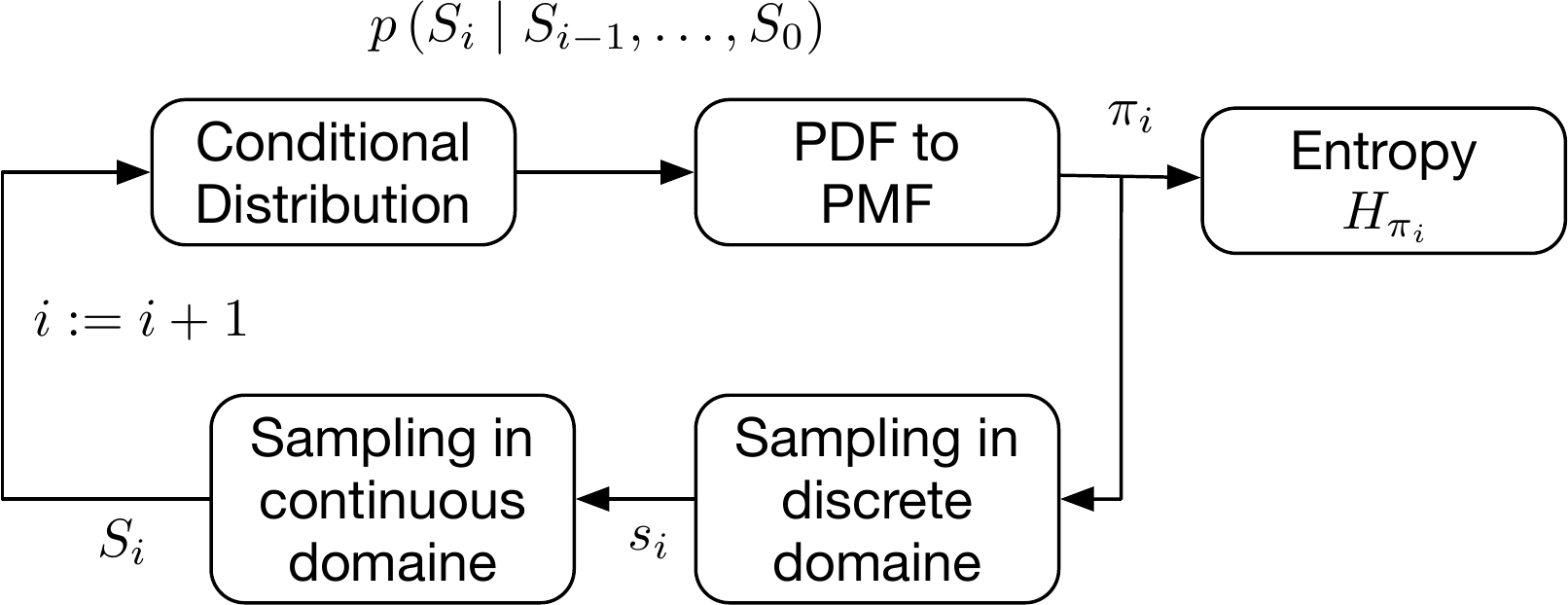}
\par\end{centering}
\caption{Sequential computation of the PMF needed to perform simulated embedding.\label{fig:Loop-over-conditional}}
\end{figure}

Note that during this step, we need to both to embed/sample on JPEG coefficients, and to sample in the continuous domain in order to be able to compute the conditional distribution using~\eqref{eq:Cond_distrib}, this is illustrated on Figure~(\ref{fig:Loop-over-conditional}).

\subsection{Entropy estimation}\label{subsec:ent}

Finally, from the probability mass function obtained in the previous
section, the binary entropy associated to the steganographic signal
for the $i^{th}$ coefficient can be computed. Given the alphabet
$\mathcal{A}=(-K,\hdots,0,\hdots,K)$, $k\in \mathbb{N}^+*$, it is defined as:
\begin{align}
H(\mathcal{A},i)=-\sum_{k\in A}\pi'_{i}(k) \,\mathrm{log_{2}}\pi'_{i}(k),\label{eq:H_estimation-1}
\end{align}
where $\pi'_{i}(k)=\pi_{i}(k)$ for $i\in  \{-K-1,\dots,K-1\}$, $\pi'_{i}(-K)=-\sum_{i=-\infty}^{i=-K}\pi_{i}(k)$ and $\pi'_{i}(K)=-\sum_{i=K}^{i=+\infty}\pi_{i}(k)$.

\subsection{Final embedding algorithm\label{subsec:Embedding}}
\begin{algorithm}[h]
- {\bf Inputs}: the cover RAW image $\mathbf{X}_p$, the payload, a secret key

- {\bf Develop} $\mathbf{X}_p$ in the DCT domain, before quantization to obtain $\mathbf{X}_d$ and in the JPEG domain to obtain $\mathbf{X}_j$;

- {\bf Divide} $\mathbf{X}_p$ into 4 macro-lattices $\Lambda_{1}$, $\Lambda_{2}$,
$\Lambda_{3}$, $\Lambda_{4}$;

- \textbf{For} each macro-lattice $\Lambda_{i}$ \textbf{do}:
\begin{itemize}
\item \textbf{For} each DCT block of $\Lambda_{i}$ \textbf{do}:
\begin{itemize}
\item Compute the covariance matrix for each set of DCT blocks (Eq.~(\ref{eq:Cov_from_RAW});
\item Compute the conditional mean vector (Eq.~(\ref{eq:conditional_mean}))
and covariance matrix (Eq.~(\ref{eq:conditional_cov})) w.r.t. the
embeddings done on the previous lattices;
\begin{itemize}
\item \textbf{For} each DCT coefficient of $\mathbf{X}_d$ \textbf{do}:
\begin{itemize}
\item Compute the conditional distribution Eq.~(\ref{eq:Cond_distrib}) given the previous embedding changes;
\item Compute the PMF $\pi_{i}(k)$, Eq.~(\ref{eq:ERF});
\item Perform the modification on $\mathbf{X}_j$ by sampling according to $\pi_{i}(k)$;
\item Sample the continuous variable related to the modification, Eq~(\ref{eq:RejecSampling});
\end{itemize}
\end{itemize}
\end{itemize}
\end{itemize}
- {\bf Return} the JPEG stego image $\mathbf{Y}_j$.
\caption{J-Cov-NS embedding scheme.\label{alg:Embedding-scheme.}}
\end{algorithm}

The resulting embedding algorithm (named J-Cov-NS) can be decomposed
into the following steps, summed up in the pseudo code presented in
Algorithm~\ref{alg:Embedding-scheme.}. The use of the key in not explicit, but it can be used to shuffle the coefficients withing each lattice. The embedded payload is such that its size matches Eq. ~\eqref{eq:H_estimation-1}.

\section{Results\label{sec:Results}}

This section presents a detailed benchmark of the embedding scheme
on JPEG images, in the cover-source switching scenario, i.e., a scenario
where the cover image comes from a higher ISO sensitivity than the
image used to generate the stego image, and where the embedding mimics
the ISO change.

\subsection{Generation of E1Base}

We evaluate the proposed embedding scheme to test on images taken by the Micro 4/3 16 MP CMOS sensor from the Z CAM E1 action camera.

\begin{figure}[H]
\begin{centering}
\includegraphics[width=0.9\columnwidth]{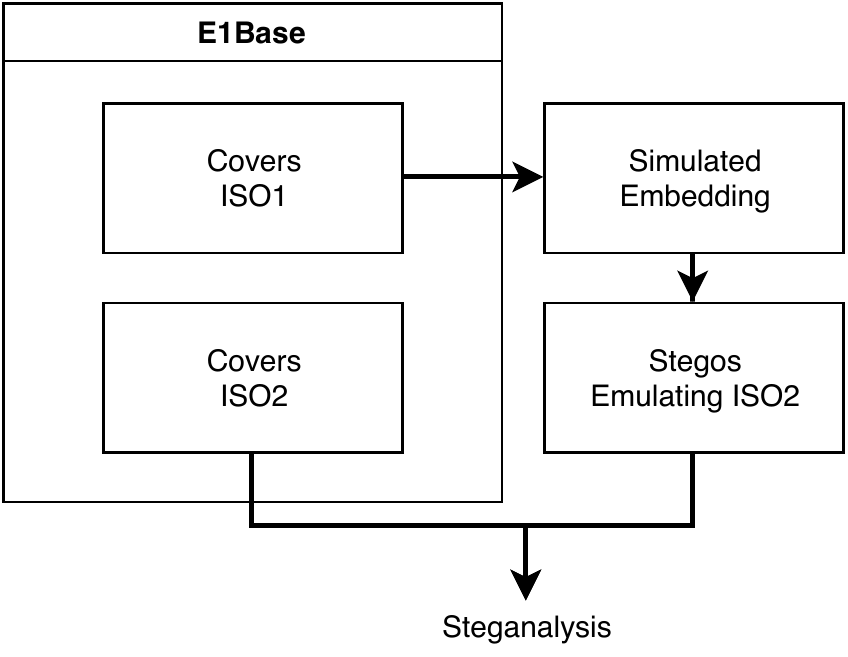}
\par\end{centering}
\caption{Steganalysis setup when benchmarking NS.\label{fig:Steganalysis-scheme-in-NS}}
\end{figure}

Note that this steganalysis setup is relatively unconventional compared to the state of the art (see Figure \ref{fig:Steganalysis-scheme-in-NS}).
This is due to the fact that the goal of the classifier here is to distinguish between cover images captured at $ISO_{2}$ from stego images coming from cover images captured at $ISO_{1}$ but emulating sensor noise captured at $ISO_{2}$.

Raw images coming from the E1 sensor are acquired with two ISO settings (ISO 100 and ISO 200) and constitute
\emph{E1Base}. 
This database can be downloaded at \href{https://gitlab.cristal.univ-lille.fr/ttaburet/e1base}{https://gitlab.cristal.univ-lille.fr/ttaburet/e1base}
and is built according to the following requirements:

- It contains an equal number of images of equivalent scenes captured at both $ISO_{1}=100$ and $ISO_{2}=200$. 
The training and testing sets have been generated from 200 Raw images (DNG format, with a 12 bits dynamic range) that have been developed and cropped without overlapping in order to provide $10,800$ images of size $512\times512$. 
This dataset has already been used under similar circumstances in \cite{denemark:hal-01687194,taburet:2019,taburet:hal-02165866}.

- A particular care has been taken in order to ensure that the only important difference between the database acquired at $ISO_{1}$ and the database acquired at $ISO_{2}$ is the sensor noise. 
In the same way as the MonoBase was acquired by a monochrome sensor \cite{bas:WIFS-16}, the average focus and average luminance are both similar between the two databases.
This step is mandatory in order to guarantee that the steganalyzer is not using semantic information to distinguish between the cover and stego datasets. 
This requirement is specific to the benchmarking process of Natural Steganography since the cover and stego images do not come from the same source in this case.

For this given database, the value used to compute the variance of the sensor noise at the photo-site level are $\left(a_{2}-a_{1}\right)=1.15$ and $\left(b_{2}-b_{1}\right)=-1150$ (the variance is set to zero whenever it is negative). A python notebook used to generate both the cover and the stego images is also downloadable here: \href{https://gitlab.cristal.univ-lille.fr/ttaburet/tifs-ns/}{https://gitlab.cristal.univ-lille.fr/ttaburet/tifs-ns/}.

Classically, {\it E1 Base} is split into two halves, 5400 pairs of images are used for training and 5400 pairs for testing.  

\subsection{Benchmark settings}

We adopt the DCTR features set~\cite{holub2015low} combined with
a low complexity linear classifier~\cite{cogranne2015ensemble} to
perform the steganalysis with the threshold set in order to minimize
the total classification error probability under equal priors, $P_{\mathrm{E}}=\min_{P_{\mathrm{FA}}}\frac{1}{2}(P_{\mathrm{FA}}+P_{\mathrm{MD}})$,
with $P_{\mathrm{FA}}$ and $P_{\mathrm{MD}}$ standing for the false-alarm
and missed-detection rates, respectively.

For comparison with the current state of the art (of side informed schemes in the JPEG domain), we embedded all images also with SI-UNIWARD with an embedding rate of 1 bit per nzAC coefficient. In this case, the steganalysis task is the classic one: try to distinguish stegos images (produced by SI-UNIWARD) from covers acquired at $ISO_{2}$.

\subsection{Comparison with other embedding strategies\label{subsec:Comparison-with-other}}

\begin{table*}[h]
\begin{centering}
\begin{tabular}{|c|c|c|c|c|c|c|c|}
\hline 
$P_{E}$ (\%) / & $H$ & \textbf{J-Cov-NS} & Pseudo & Covariance & Independent & Intra-block & SI-Uniward \cite{holub2014universal}\tabularnewline
JPEG QF & (bpnzAC) & & embedding \eqref{eq:Stego_noise_CSS} & scaling \cite{taburet:2019} & embedding \cite{denemark:hal-01687194} & correlations only & 1 bpnzAC\tabularnewline
\hline 
100 & 2.0 &\textbf{42.9} & 40.2 & 13.9 & 0.0 & 0.0 & 0.0\tabularnewline
\hline 
95 & 2.2 &\textbf{41.2} & 40.9 & 30.3 & 0.5 & 0.2 & 0.4\tabularnewline
\hline 
85 & 2.4 &\textbf{41.2} & 41.9 & 39.8 & 10.8 & 15.8 & 12.3\tabularnewline
\hline 
75 & 7.0 &\textbf{41.6} & 41.3 & 40.4 & 27.0 & 25.2 & 24.8\tabularnewline
\hline 
\end{tabular}
\par\end{centering}
\medskip{}

\caption{Empirical security ($P_{\mathrm{E}}$ in \%) and average embedding capacity ($H$) for different quality
factors and embedding strategies on E1Base. DCTR features combined
with regularized linear classifier are used for steganalysis.\label{tab:Empirical-security-}}
\end{table*}

Table~\ref{tab:Empirical-security-} compares the proposed embedding scheme for different JPEG QF with other embedding strategies which are: 
\begin{itemize}
\item Pseudo embedding in the photo-site domain, i.e. using Eq.~(\ref{eq:pseudo-emb}), and applying the process depicted in the top raw of Figure~\ref{fig:Development-of-a-1}, 
\item Estimating the empirical covariance matrix from a stationary signal and scaling it according to the average RGB values of the raw image, which is one solution to circumvent the explicit calculus of the covariance matrix~\cite{taburet:2019}, 
\item Embedding without taking into account correlations between DCT coefficients, this is performed by computing an empirical histogram of each DCT mode estimated after multiple embeddings and Monte-Carlo simulations~\cite{denemark:hal-01687194}, 
\item Embedding taking into account only intra-block correlations, this is performed by using only the computation of the intra-block covariance matrix, no inter-block correlations are consequently considered here, 
\item SI-UNIWARD~\cite{holub2014universal}, one state of the art embedding scheme in the JPEG domain which use side-informed embedding from the RAW image.
\end{itemize}

We can notice that computing the covariance matrix for each DCT block enables us to achieve about the same practical security than pseudo-embedding. Contrary to the previous scheme proposed in~\cite{taburet:2019}, which relies on an approximation of the covariance matrix using a scaling factor dependent on the RGB values of each block, J-Cov-NS
does not exhibit any security loss for high QFs. The comparison with independent embedding, which offers good practical security for monochrome sensors, highlights the fact that the latter scheme is not adapted to color sensors, and that it is extremely important to take into account correlations between DCT coefficients, especially for high
QFs. Note also that if only the intra-block correlations are taken into account, the embedding scheme still remains highly detectable.
Finally, the comparison with SI-UNIWARD shows that this state-of the art scheme is not secure for very high embedding rates (1 bit pnzAC coefficient here). This is not surprising since SI-UNIWARD does not rely on cover-source switching and does not use all the information provided by the development pipeline.

\subsection{Evaluation for other steganalysis strategies\label{subsec:Evaluation-with-steganalysis}}

We also evaluated J-Cov-NS w.r.t to other steganalysis strategies dedicated to JPEG images. To this end, we performed steganalysis using another JPEG feature sets based on resudials extracted using Gabor filters (GFR, see~\cite{song2015steganalysis}) and also using the non-linear ensemble classifier~\cite{kodovsky2012ensemble} for different JPEG QF. Results are presented in Table~\ref{tab:GFRandEC} and shows that both strategies are equivalent with the former one, with a slight advantage on GFR over DCTR features (-1\% to -3\%). Note however that GRF features have higher dimensionality ($17.10^3$ vs $8.10^3$) and are longer to extract. The use of the ensemble classifier enables also gain reduce the detectability, but by a small margin of maximum 1\%, together with a computational cost of about one order of magnitude. %We hypothesize that (i) GFR are not adequate to model the photonic noise and (ii) that ensemble classifier, which are composed of a set of weak classifiers, are in this case not appropriate because each base learner is too weak.

Since steganalysis based on deep neural network offer the opportunity to automatically extract relevant features regardless of the embedding scheme, we also benchmark J-Cov-Net w.r.t. SRNet, one state of the art network in spatial or JPEG steganalysis~\cite{boroumand2018deep}. The network was trained using mini-batches of 32 $512\times 512$ images (16 covers and 16 stego) using Nvidia GPU Quadro P6000 (24 GB of memory), the learning rate was is initially set to $10^{-3}$ and decreases by 10\% each 5000 iterations. The sizes of the training set is 4000 pairs (augmented using rotations and flipping transforms), 1000 pairs are used for validation in order to select the best trained network, and the rest for testing. The results presented in table~\ref{tab:SRNet} are obtained after convergence is reached, i.e. after 100 000 iterations. We can notice that DNN based steganalysis enables to increase the performances in detectability by about 10\% w.r.t. to DCTR combined with the low complexity linear classifier. With more than 30\% of average error rate, this does not jeopardize the detectability of the presented scheme though. Note also that this improved detectability can be due to the fact that the automatic feature extraction provided by the convolutional layers of SRNet succeeds to catch possible slight general content discrepancies between images of E1Base acquired at ISO 100 and 200.

\begin{table}[h!]
\begin{centering}
\begin{tabular}{|c|c|c|c|c|}
%\cline{1-1} \cline{3-6} \cline{4-6} \cline{5-6} \cline{6-6} \cline{8-8} 
\hline
QF / & \multicolumn{2}{c|}{Linear Classifier} & \multicolumn{2}{c|}{Ensemble Classifier} \tabularnewline
%\cline{3-6} \cline{4-6} \cline{5-6} \cline{6-6} 
$P_{E}$ (\%) & DCTR & GFR & DCTR & GFR \tabularnewline
%\cline{1-1} \cline{3-6} \cline{4-6} \cline{5-6} \cline{6-6} \cline{8-8} 
\hline
\hline %
100 & 42.9 & 40.3 & 40.8 & 39.6 \tabularnewline
\hline
%\cline{1-1} \cline{3-6} \cline{4-6} \cline{5-6} \cline{6-6} \cline{8-8} 
95 & 41.2 & 39.2 & 	41.3 & 38.4 \tabularnewline
%\cline{1-1} \cline{3-6} \cline{4-6} \cline{5-6} \cline{6-6} \cline{8-8} 
\hline
85 & 41.2 & 39.1 & 41.0 & 38.1 \tabularnewline
\hline

%\cline{1-1} \cline{3-6} \cline{4-6} \cline{5-6} \cline{6-6} \cline{8-8} 
75 & 41.6 & 40.3 & 41.4 & 39.1 \tabularnewline
%\cline{1-1} \cline{3-6} \cline{4-6} \cline{5-6} \cline{6-6} \cline{8-8} 
\hline
\end{tabular}
\par\end{centering}
\medskip{}

\caption{Practical security of J-Cov-NS for other steganalysis strategies: DCTR and GFR features sets using the Linear Classifier and the Ensemble Classifier.}\label{tab:GFRandEC}
\end{table}

\begin{table}[h!]
\begin{centering}
\begin{tabular}{|c|c|c|c|}
\hline
QF & 100 & 95 & 75\\
\hline
$P_E\,(\%)$ & 37.4 & 31.2 & 35.0\\
\hline
\end{tabular}
\par\end{centering}
\medskip{}

\caption{Practical security of J-Cov-NS against SRNet.\label{tab:SRNet}}
\end{table}

\subsection{Embedding capacity\label{subsec:Payload-as-a}}

\begin{figure}[th]
\begin{centering}
\subfloat[\label{fig:H_K_coeff}]{\begin{centering}
\includegraphics[width=0.49\columnwidth]{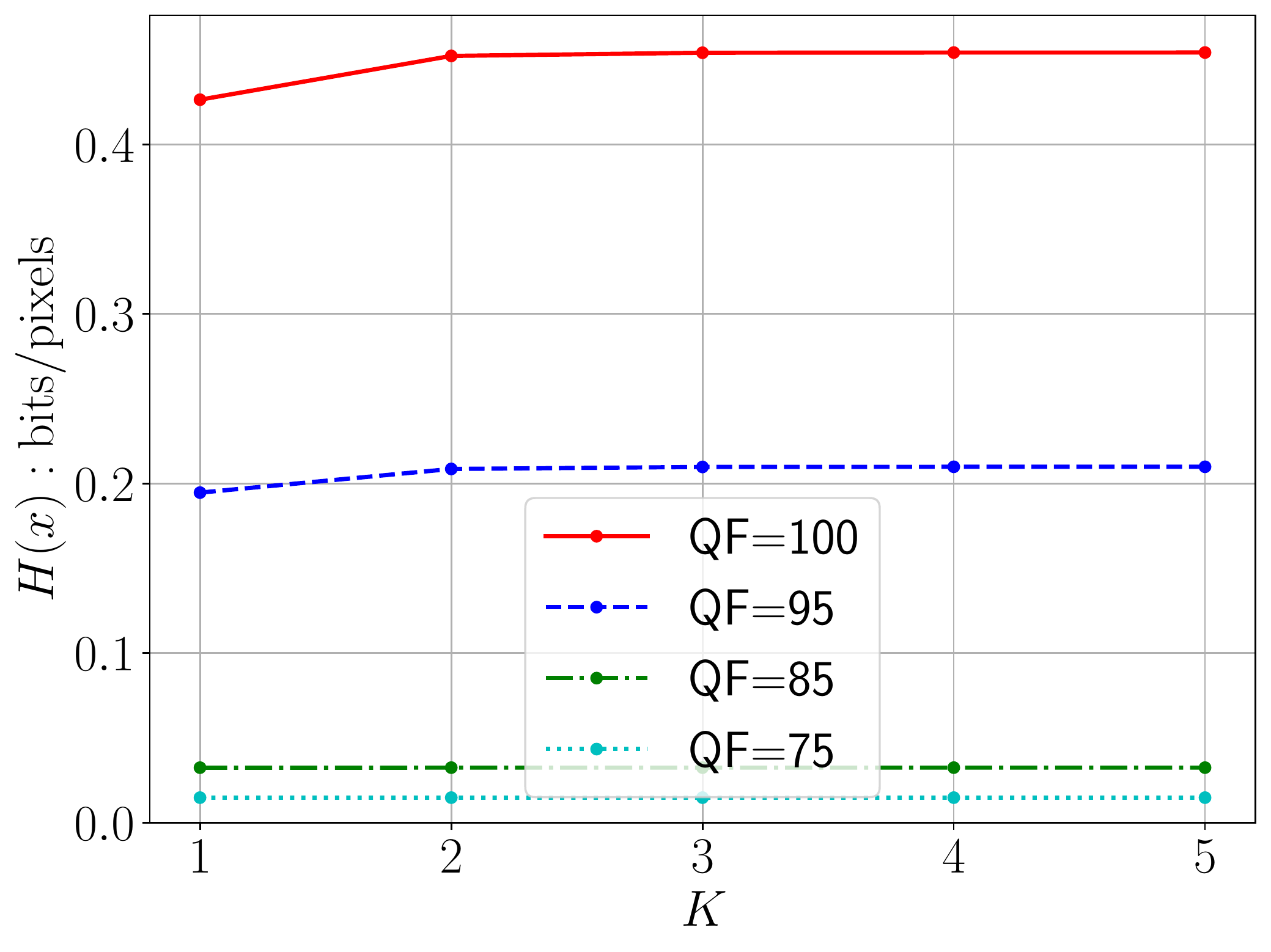}
\par\end{centering}
}\subfloat[\label{fig:H_K_nzAC}]{\begin{centering}
\includegraphics[width=0.49\columnwidth]{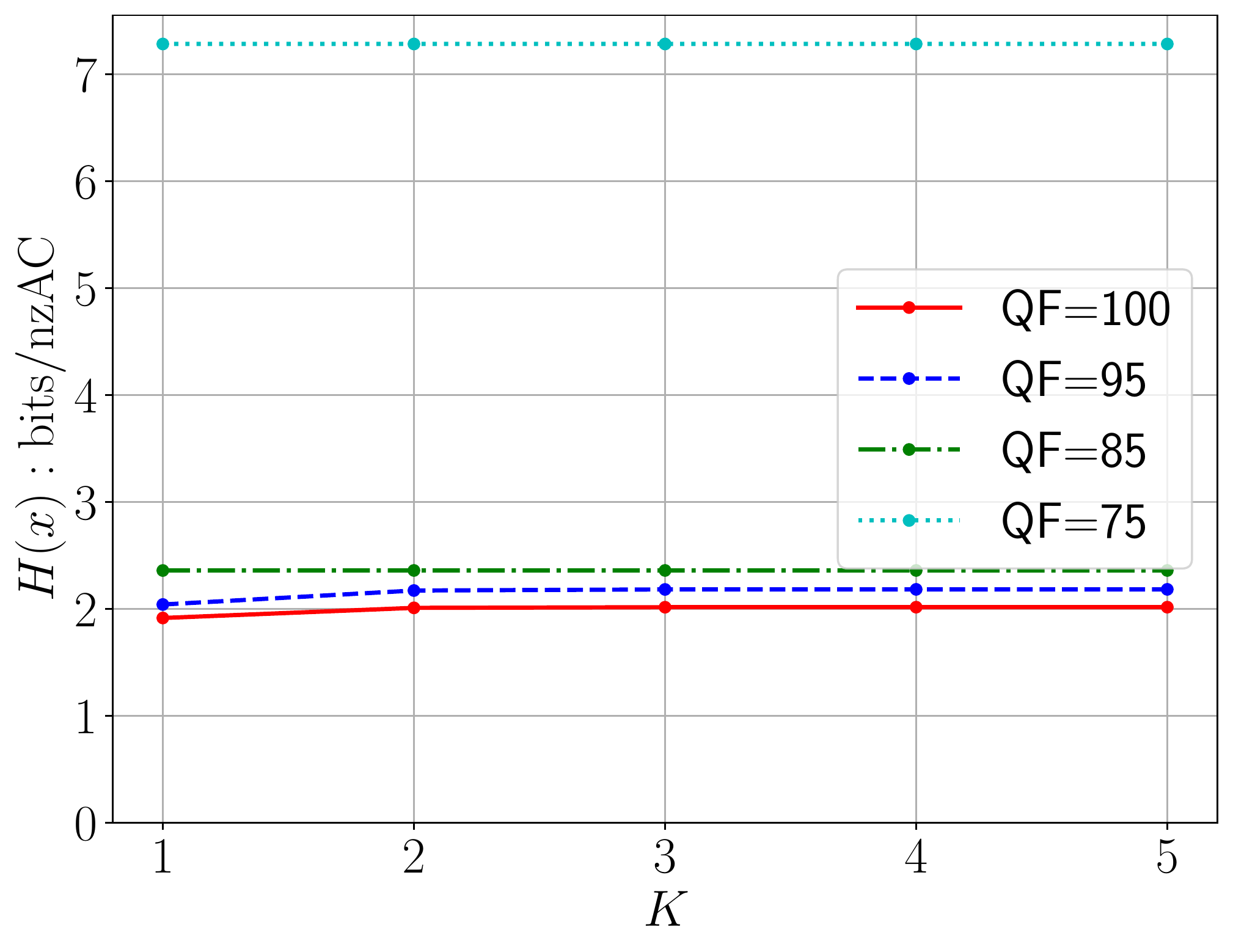}
\par\end{centering}
}
\par\end{centering}
\begin{centering}
\subfloat[\label{fig:H_K_coeff-1}]{\begin{centering}
\includegraphics[width=0.49\columnwidth]{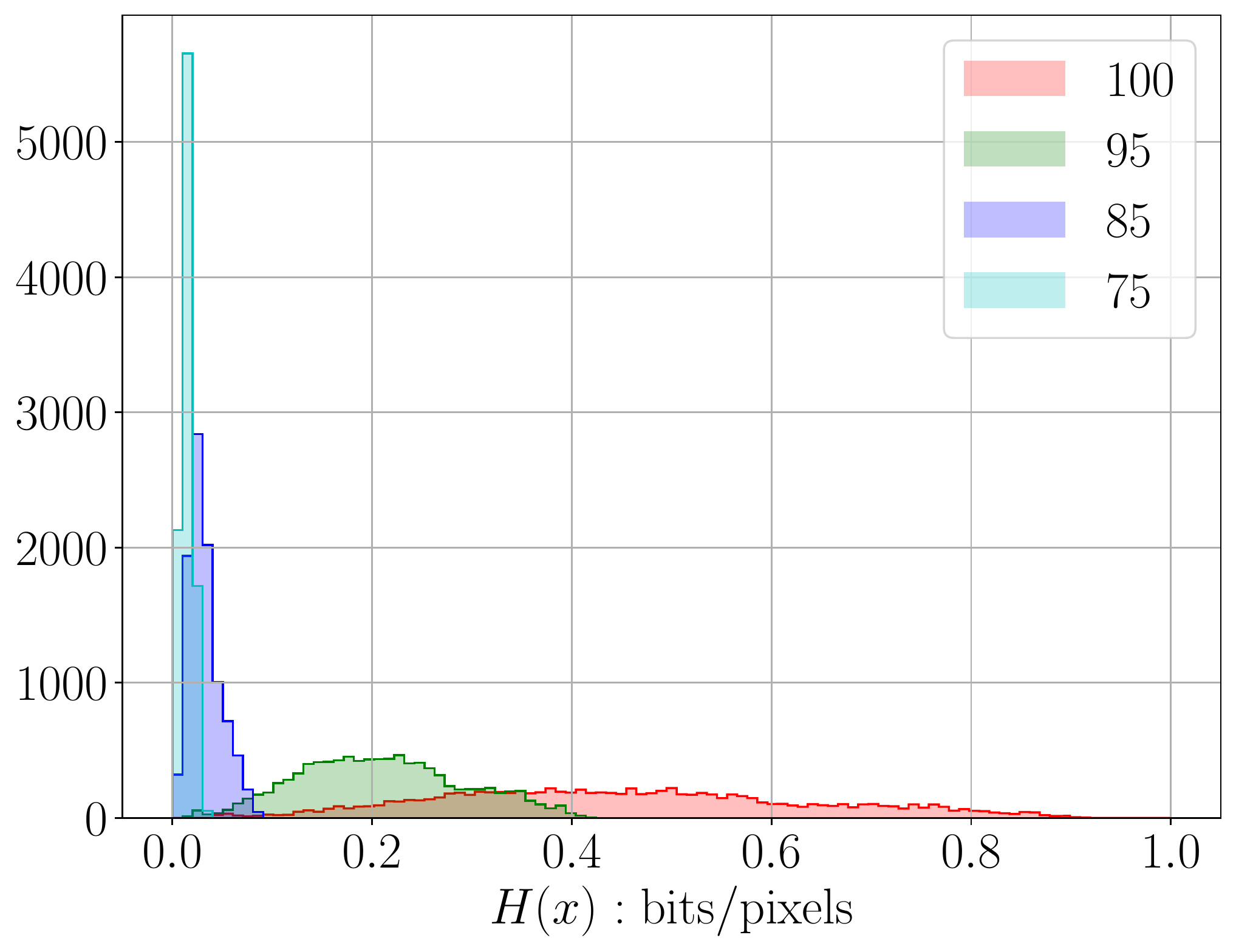}
\par\end{centering}
}\subfloat[\label{fig:H_K_nzAC-1}]{\begin{centering}
\includegraphics[width=0.49\columnwidth]{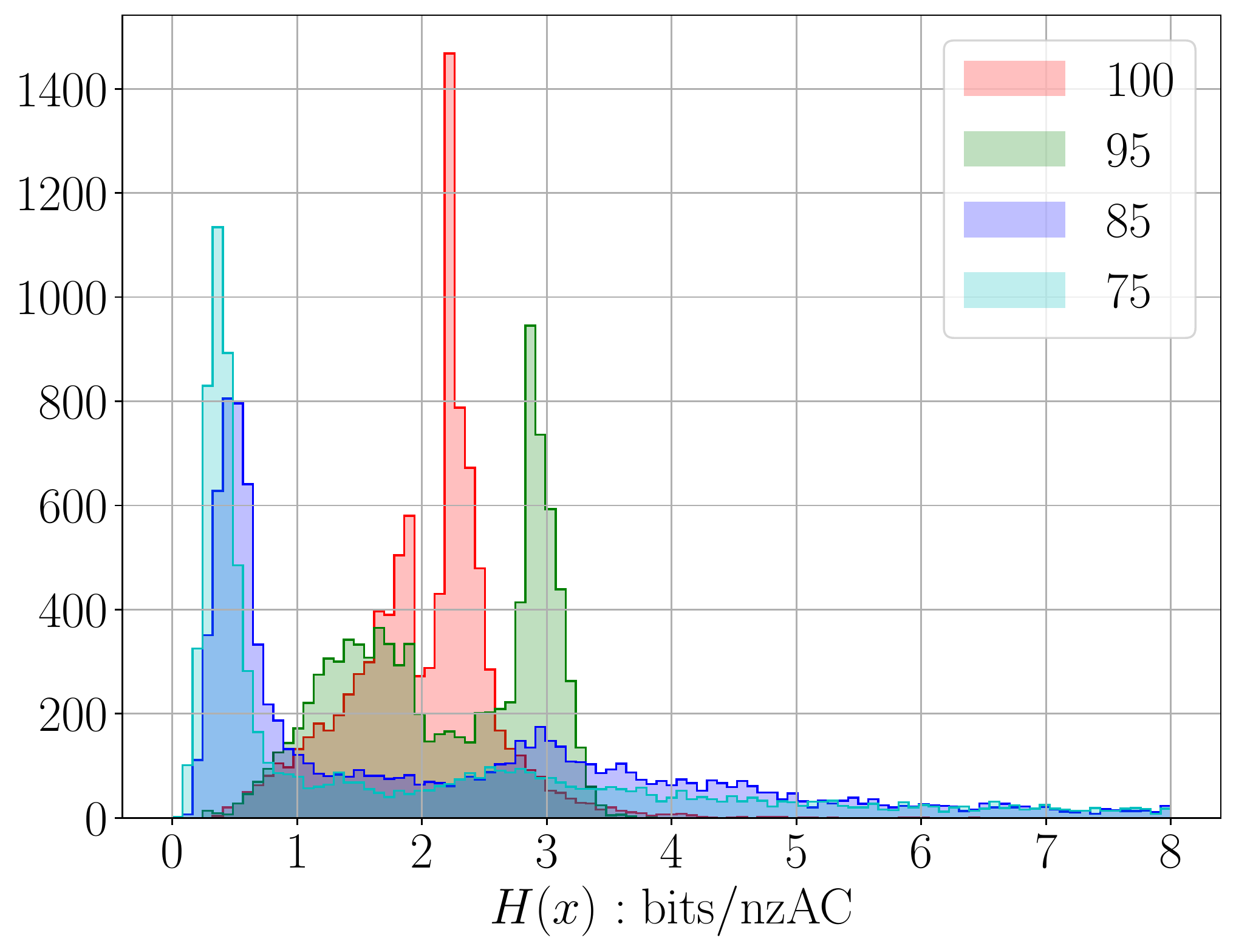}
\par\end{centering}
}
\par\end{centering}
\caption{Average entropy $H$ (bits) of J-Cov-NS over the database (a) per
pixel, (b) per nzAC as a function of $K$ for different JPEG QFs.
Histograms of $H$ (bits) across images for different QFs in (c) per
pixel, (d) per nzAC.}
\end{figure}

In this section, we investigate the distribution of the embedding
capacity through the whole E1Base database, and compute its average
value for JPEG QFs $75$, $85$, $95$, and $100$ and for different
alphabet sizes. Thus, we estimate the entropy for each $512\times512$
image, compute the proportion of nzAC and obtain $H_{bits/pixels}$
and $H_{bits/nzAC}$ as a function of the of the chosen alphabet size
for each QF. Figure~\ref{fig:H_K_coeff} and Figure~\ref{fig:H_K_nzAC}
illustrate, respectively, the evolution of $H_{bits/pixels}$ and
$H_{bits/nzAC}$ when the size of the alphabet for insertion increases
from $\left[\begin{array}{ccc}
-1 & 0 & +1\end{array}\right]$ to $\left[\begin{array}{ccc}
-5 & \ldots & +5\end{array}\right]$.

The average embedding capacity in bits per nzAC is relatively high,
around 2 bits pnzAC for JPEG QF$\in\{95,100\}$ and over 7 bits pnzAC
for $QF\in\{75,85\}$. The alphabet size has a minor impact on the
capacity. However, $\mathrm{QF}\in\{75,85\}$ highlights an exotic
case, since on the one hand the embedding is concentrated on the DC
coefficients, and on the other hand there are only few nzAC coefficients
at $QF\in\{75,85\}$. For example, given a $512\times512$ image with
an average embedding rate of $1$ bit per DC coefficient and having
only $100$ non-zero AC coefficients, this image has a total embedding
rate of $40.96$ bits per nzAC!

\begin{figure}[h]
\begin{centering}
\subfloat[$QF=100$]{\begin{centering}
\includegraphics[width=0.48\columnwidth]{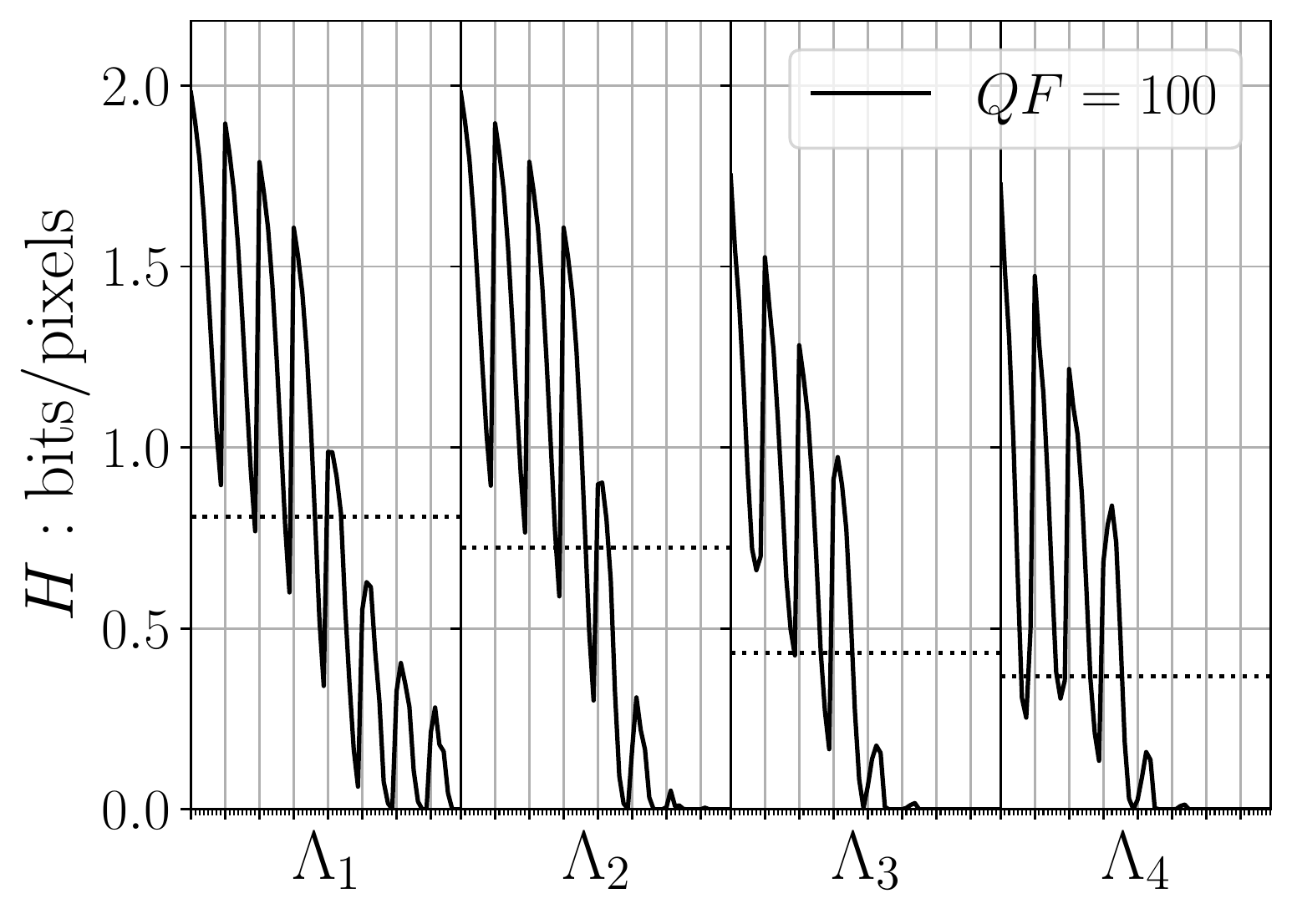}
\par\end{centering}
}\subfloat[$QF=95$]{\begin{centering}
\includegraphics[width=0.49\columnwidth]{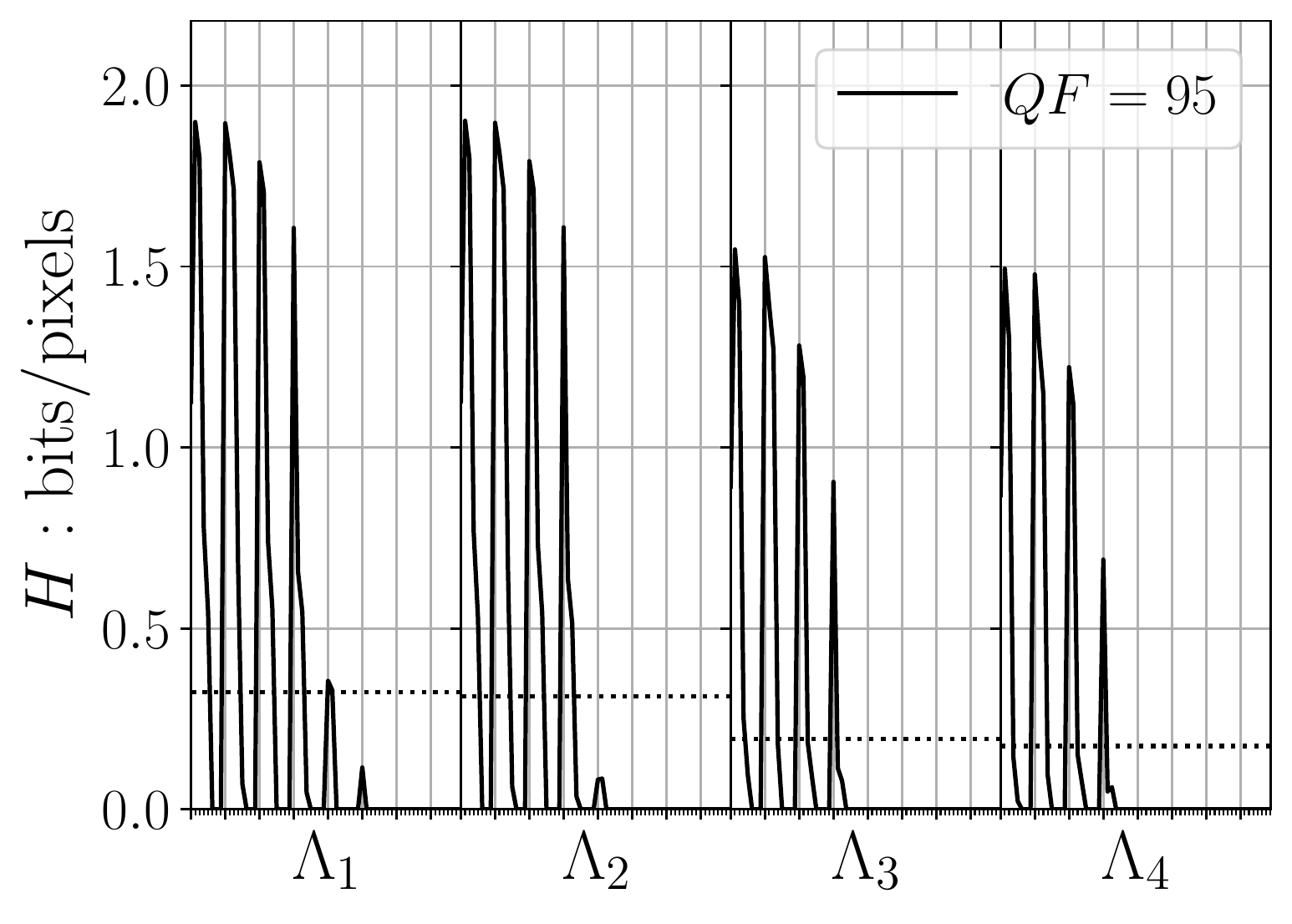}
\par\end{centering}
}
\par\end{centering}
\centering{}\caption{Evolution of the embedding rates computed from an i.i.d. Gaussian
RAW image for each DCT mode and each sub-lattice for different JPEG QFs. 
Row scan is used within each sub-lattice. 
Dotted lines denote the average embedding rate within each sub-lattice.\label{fig:Evolution-Sublattices}}
\end{figure}

Figure~(\ref{fig:Evolution-Sublattices}) shows the embedding capacity
computed on a synthetic constant cover RAW image for each DCT coefficient
on the four lattices $\Lambda_{1}$, $\Lambda_{2}$, $\Lambda_{3}$,
and $\Lambda_{4}$ at $\mathrm{QF}=100$ and $\mathrm{QF}=95$. Within
each block, row scan is used. Two remarks can be drawn:
\begin{enumerate}
\item the capacity decreases w.r.t. the coefficient frequency, this is due
to demosaicking and the fact that the stego signal is mainly encoded
by low frequency components. For $\mathrm{QF}=95$, this is also due
to the fact that the quantization steps are larger for high frequencies.
\item the capacity decreases w.r.t. the lattice index, with an average value
at $\mathrm{QF}=100$ of 0.8 bpp for $\Lambda_{1}$ to 0.4 bpp for
$\Lambda_{4}$. This is because conditioning reduces the entropy of
a random variable~\cite{cover1991elements}. At $\mathrm{QF}=100$,
where the quantization is the same for each DCT mode, this is particularly
noticeable by examining the entropy of the last 8 coefficients of
each block, which are up to 0.3 bpp for $\Lambda_{1}$ but, due to
conditioning, are reduced to zero for $\Lambda_{4}$.
\end{enumerate}

\subsection{Impact of the alphabet size}

The impact of the alphabet size $(2K+1)$ on the implementation of J-Cov-NS is presented in Table~(\ref{tab:Practical-security-w.r.t.}) for different JPEG QF. 
We can notice that ternary embedding ($K=1$) is associated with a very detectable implementation for $\mathrm{QF}=95$ and $\mathrm{QF}=100$. 
This is due to the fact that the truncation of the modification changes alters considerably the distribution of the stego signal which cannot mimic anymore the ISO switch for small quantization steps. 
On the other hand, heptary embedding offers detectability comparable to that of an infinite alphabet for $\mathrm{QF}=95$ and should be used for true embedding combined with multi-layer STC in this case. We can also notice that for $\mathrm{QF}\leq85$ ternary embedding offers already the same practical security than pentary embedding.

\begin{table}[H]
\begin{centering}
\begin{tabular}{|c|c|c|c|c|}
\hline 
QF / & \multirow{2}{*}{$K=1$} & \multirow{2}{*}{$K=2$} & \multirow{2}{*}{$K=3$} & \multirow{2}{*}{$K=5$}\tabularnewline
$P_{E}$ in \% &  &  &  & \tabularnewline
\hline 
\hline 
100 & 1.0 & 12.9 & 28.7 & 40.4\tabularnewline
\hline 
95 & 3.5 & 23.6 & 39.3 & 40.9\tabularnewline
\hline 
85 & 39.8 & 39.8 & 39.8 & 41.8\tabularnewline
\hline 
75 & 40.4 & 40.4 & 40.4 & 41.2\tabularnewline
\hline 
\end{tabular}
\par\end{centering}
\medskip{}

\caption{Practical security of J-Cov-NS w.r.t. alphabet size and different $QF$.\label{tab:Practical-security-w.r.t.}}
\end{table}

\subsection{Complexity consideration}

This embedding algorithm is computationally expensive since the complexity of computing the conditional distribution increases as the complexity of the Cholesky decomposition of the covariance matrix, i.e., as $\mathcal{O}(n^{3})$ where $n\leq i\times64$, where $i=1$ for $\Lambda_{1}$, $i=5$ for $\Lambda_{2}$ and $\Lambda_{3}$, and $i=9$ for $\Lambda_{4}$ (see Figure~\ref{fig:Lattices}). 
On a 3.5 GHz Intel Core i7, our python implementation of simulated embedding is executed at 4000 block/s for blocks belonging to $\Lambda_{1}$, 30 blocks/s for $\Lambda_{2}$, 30 blocks/s for $\Lambda_{3}$ and 10 blocks/s for $\Lambda_{4}$.
A $512\times512$ stego is generated in approximately 171s without using hyper-threading.

\section{Conclusions and perspectives}

This paper draws important conclusions both in image processing and
image steganography.

By deriving the covariance matrix of the random vector of stego signal components in the DCT domain, we have shown that for this basic development pipeline there are medium range correlations between DCT coefficients, and that for a given coefficient, it is correlated with the coefficients belonging to the same blocks, but also with the coefficients belonging to 8-connected blocks.
Previous works on the estimation of the covariance matrix were conducted for denoising applications using non-local Bayesian estimation~\cite{lebrun2015noise}, but to the best of our knowledge, it is the first time that an analytical expression of the covariance matrix is derived in the DCT domain (i.e. Eq.~\eqref{eq:Vect_DCT_2D_from_RAW} and~\eqref{eq:Cov_from_RAW}), exhibiting intra-block and inter-block correlations.

The derivation of the covariance matrix enables to generate a stego signal that mimics the photonic noise in the DCT domain and consequently to achieve high practical security ($P_{E}\geq40\%$ for DCTR features set) while reaching high capacity ($>2$ bpnzAC). 
In order to preserve the joint Gaussian distribution after embedding in the quantized DCT domain, the J-Cov-NS embedding scheme needs to use a large number of lattices ($4\times64$) where conditional probability mass functions are derived for each lattice. 
Our experimental analysis shows that for high JPEG QF, being able to perform conditioning is essential to achieve high practical security. A similar synchronization strategy was also adopted for adaptive schemes using empirical costs in~\cite{li2018defining} and~\cite{taburet2020jpeg}.

Our future works will focus on non-linear developments, which may decrease the security of the scheme in a highly non linear case (see~\cite{taburet:2019}), and on designing a similar scheme for color stego images, which mean that we will need to model correlations between color channels.

\section*{Acknowledgments}
The authors would like to thank Solène Bernard (from CRIStAL Research Center, Lille) for running the experiments using SRNet.

\bibliographystyle{plain}
\bibliography{totaleBibDesk}

\end{document}